 \documentclass{emulateapj}
  \usepackage{epsfig}
 \usepackage{subfig}
\shorttitle{Spectral Evolution of GRB X-Ray Tails} \shortauthors{Zhang et al. }
\begin{document}

\title{A Comprehensive Analysis of {\em Swift} XRT Data. I. \\ Apparent Spectral Evolution of Gamma-Ray Burst X-ray tails}
\author{Bin-Bin Zhang\altaffilmark{1,2}, En-Wei Liang \altaffilmark{1,3},
Bing Zhang\altaffilmark{1}
} \altaffiltext{1}{Department of Physics and Astronomy, University of Nevada, Las Vegas,
NV 89154, USA; zbb@physics.unlv.edu; lew@physics.unlv.edu; bzhang@physics.unlv.edu}
\altaffiltext{2}{National Astronomical Observatories/Yunnan Observatory, CAS, Kunming
650011, China}
\altaffiltext{4}{Department of Physics, Guangxi University,
Nanning 530004, China}

\begin{abstract}
An early steep decay component following the prompt GRBs is commonly observed in {\em Swift} XRT light curves, which is regarded as the tail emission of the prompt gamma-rays. Prompted by the observed strong spectral evolution in the tails of GRBs 060218 and 060614, we present a systematic time-resolved spectral analysis for the {\em Swift} GRB tails detected between  2005 February and  2007 January.  We select a sample of 44 tails that are bright enough to perform time-resolved spectral analyses. Among them 11 tails are smooth and without superimposing significant flares, and their spectra have no significant temporal evolution. We suggest that these tails are dominated by the curvature effect of the prompt gamma-rays due to delay of propagation of photons from large angles with respect to the line of sight . More interestingly, 33 tails show clear hard-to-soft spectral evolution, with 16 of them being smooth tails directly following the prompt GRBs,while the others being superimposed with large flares. We focus on the 16 clean, smooth tails and consider three toy models to interpret the spectral evolution. The curvature effect of a structured jet and a model invoking superposition of the curvature effect tail and a putative underlying soft emission component cannot explain all the data. The third model, which invokes an evolving exponential spectrum, seems to reproduce both the lightcurve and the spectral evolution of all the bursts, including GRBs 060218 and 060614. More detailed physical models are called for to understand the apparent evolution effect. 
\end{abstract}


\section{INTRODUCTION\label{sec:intro}}
The extensive observations of gammar-ray bursts (GRBs) suggest that most of the broadband, power-law decaying
afterglows are from external shocks as the fireball is decelerated by the ambient medium
(M\'esz\'aros \& Rees 1997a; Sari et al. 1998). The prompt gamma rays and the erratic
X-ray flares after the GRB phase (Burrows et al. 2005),  are instead of internal origin,
likely from internal shocks (Rees \& M\'esz\'aros 1994, see Zhang et al. 2006 for
detailed discussion)\footnote{Most recently Ghisellini et al. (2007) suggested that most
power-law decaying X-ray afterglows that show a shallow-to-normal decay transition are
``late prompt emission'' that is also of internal origin. The fact that most of the X-ray
afterglows in the ``normal'' decay phase satisfy the well-known ``closure relation'' for
the external shocks (Zhang et al. 2007a; see also the second paper in this series, Liang
et al. 2007, Paper II), however, suggests that this is not demanded for most bursts. GRB
0070110, on the other hand, displays a flat X-ray emission episode followed by a rapid
decay. This likely suggests an internal origin of the flat X-ray emission episode at
least for some bursts (Troja et al. 2007).}. The direct evidence for the distinct
internal origin of prompt gamma-rays and X-ray flares is the steep decay tails following
the prompt emission and the flares (Tagliaferri et al. 2005; Nousek et al. 2006; O'Brien
et al. 2006), which could be generally interpreted as the so-called ``curvature effect''
due to the delay of propagation of photons from high latitudes with respect to the line
of sight (Fenimore et al. 1996; Kumar \& Panaitescu 2000; Qin et al. 2004; Dermer 2004;
Zhang et al. 2006; Liang et al. 2006a). This clean picture is somewhat ``ruined'' by some
recent observations with {\em Swift}. A strong spectral evolution has been observed in
the tails of two peculiar GRBs: 060218 (Campana et al. 2006; Ghisellini et al. 2006) and
060614 (Gehrels et al. 2006; Zhang et al. 2007b; Mangano et al. 2007), which is not
directly expected from the curvature effect model. This suggests that there might be
unrevealed emission components in the early afterglow phase. This motivates us to perform
a systematic data analysis for both light curves and their spectral evolution of the GRB
tails observed by {\em Swift}/XRT. Our data reduction and sample selection are delineated
in \S2. The light curves and spectral evolutions are presented in \S3. In \S4, we
discuss three models, and identify an empirical model to interpret the
data. Our conclusions are summarized in \S5.

\section{DATA REDUCTION AND SAMPLE SELECTION\label{sec:data}}
The X-ray data are taken from the Swift data archive. We develop a script to
automatically download and maintain all the Swift X-Ray Telescope (XRT) data. The Heasoft
packages, including {\em Xspec}, {\em Xselect}, {\em Ximage}, and Swift data analysis
tools, are used for the data reduction. We develop a set of IDL codes to automatically
process the XRT data. The procedure is described as follows.

First, run the XRT tool {\em xrtpipeline} (Version 0.10.6) to reproduce the XRT clean
event data, which have been screened with some correction effects (e.g. bad or hot pixels
identifications, correct Housekeeping exposure times, etc.). The latest calibration data
files (CALDB, released on Dec 06,2006) are used.

Second, a time filter for the time-resolved spectral analysis is automatically performed. We initially divide the time series of XRT data into $n$ (normally 30) equal segments in log-scale. Generally, these segments are not the {\rm real} time intervals to perform the spectral analysis because they may not have enough spectral bins to perform spectral fitting. A real time interval for our spectral analysis should satisfy two criteria, i.e., the spectral bins\footnote{We re-group the spectra using {\em grppha} in order to ensure a minimum of 20 counts per spectral bin} in the time interval should be greater than 10, and the reduced $\chi^2$ should be around unity. If one temporal segment does not satisfy our criteria, we combine the next time segment until the merged segment meets our criteria. With this procedure, we create a time filter array to perform time-resolved spectral analyses.

Third, make pile-up correction and exposure correction for each time interval. The
pile-up correction is performed with the same methods as discussed in Romano et al.
(2006) (for the Window Timing [WT] mode data) and Vaughan et al. (2006) (for the Photon
Counting [PC] mode data). Both the source and the background regions are annuluses (for
PC) or rectangular annuluses (for WT) . For different time intervals,  the inner radius
of the (rectangular) annulus are dynamically determined by adjusting the inner radius of
the annulus through fitting the source brightness profiles with the King's point source
function (for PC) or determined by the photo flux using the method described in  Romano
et al. (2006) (for WT). If the pile-up effect is not significant, the source regions are
in shape of a circle with radius $R=20$ pixels (for PC) or of a $40\times 20$ pixel$^2$
rectangle (for WT) centered at the bursts' positions. The background region has the
same size as the source region, but is 20 pixels away from the source
region. The exposure correction is made with an exposure map created by the XRT tools
{\em xrtexpomap} for this given time interval.

Fourth, derive the corrected and background-subtracted spectrum and light curve for each
time interval. The signal-to-noise ratio is normally 3, but we do not rigidly fix it
to this value. Instead we adjust it if needed according to the source brightness at a
given time interval.

Fifth, fit the spectrum in each time interval and convert the light curve in count rate
to energy flux. The spectral fitting model is a simple power-law combined with the
absorptions of both our Galaxy and the GRB host galaxy, $wabs^{\rm Gal}\times zwabs^{\rm
host}\times powerlaw$ (for bursts with known redshifts) or $wabs^{\rm Gal}\times
wabs^{\rm host}\times powerlaw$ (for bursts whose redshifts are unknown), except for
GRB060218, for which a black body component is added to the fitting model,
$wabs^{\rm Gal}\times wabs^{\rm
host}\times(powerlaw+bbodyrad)$ (Campana et al. 2006)\footnote{We fix the parameters of
the black body component to the same values as in Campana et al. (2006) (see also
http://www.brera.inaf.it/utenti/campana/060218/060218.html). Please note that the XRT
light curve of the first orbit is dominated by the black body component 2000 seconds
since the GRB trigger. Therefore, the non-thermal emission in the first orbit is
considered only for those before 2000 seconds since the GRB trigger.}. The $nH^{\rm Gal}$
value is taken from Dickey \& Lockman (1990), while the $nH^{\rm host}$ is taken as a
free parameter. We do not consider the variation of $nH^{\rm host}$ within a burst and
fix this value to that derived from the time-integrated spectral fitting. With the
spectrum in this time interval, we convert the photon flux to the energy flux.

We perform time-resolved spectral analyses with our code for all the Swift GRBs detected
from Feb. 2005 to Jan. 2007, if their XRT data are available. We find that the X-rays of
most GRBs are not bright enough to make time-resolved spectral analyses, i.e., only
time-integrated spectra are derived. In this paper we focus on the spectral evolution of
GRB tails. Therefore, our sample includes only those bursts that have bright GRB tails.
All the tails studied have decay slopes $\alpha < -2$, and the peak energy fluxes in the
tails are generally greater than  $10^{-9}$ erg cm$^{-2}$ s$^{-1}$.
Some GRB tails are superimposed with significant flares. Although it is difficult to
remove the contamination of the flares, we nonetheless include these bursts as well
in the sample. Our sample include 44 bursts altogether. Their lightcurves and time-dependent
spectral indices are displayed in Figs. 1-3.

\section{RESULTS OF TIME-RESOLVED SPECTRAL ANALYSES\label{sec:result}}
The light curves and spectral index evolutions of the GRB tails in our sample are shown
in Figs. 1-3. For each burst, the upper panel shows the light curve and the
lower panel shows the evolution of the spectral index $\beta$ ($\beta=\Gamma-1$, where
$\Gamma$ is the photon index in the simple power-law model $N(E)\propto \nu^{-\Gamma}$).
The horizontal error bars in the lower panel mark the time intervals. For the purpose
of studying tails in detail, we zoom in the time intervals that enclose the tails.
In order to compare the spectral behaviors of the shallow decay phase following the
GRB tail, we also show the light curves and spectral indices of the shallow decay phase, if they were detected.

Shown in Fig.1 are those tails (Group A) whose light curves are smooth and free of
significant flare contamination, and whose spectra show no significant evolution. The
spectral indices of the shallow decay segment following these tails are roughly
consistent with those of the tails. Figure 2 displays those tails (Group B) that have
clear hard-to-soft spectral evolution\footnote{We measure the spectral evolution of these bursts with $\beta_{\rm XRT}\propto {\kappa}\log t$, and the $\kappa$ values of these bursts are greater than 1.}, but without significant flares (although some flickering has
been seen in some of these tails). The spectral evolution of these tails should be
dominated by the properties of the tails themselves, and this group of tails are the
focus of our detailed modeling in \S4. In contrast to the tails shown in Fig. 1, the
spectra of the shallow decay components following these tails are dramatically harder
than the spectra at the end of the tails. This indicates that the tails and the shallow
decay components of these bursts have different physical origin.

The rest of the GRBs (about 1/3) in our sample show those tails (Group C) that are
superimposed with significant
X-ray flares. In most of these tails, strong spectral evolutions are also observed.
These bursts are shown in Fig.3. Since the spectral behaviors may be complicated by
the contributions from both the tails and the flares, modeling these tails is no longer
straightforward, and we only present the data in Fig.3.

\section{MODELING THE TAILS: AN EMPIRICAL SPECTRAL EVOLUTION MODEL}
The physical origin of the GRB tails is still uncertain. In our sample, one-fourth
of the tails do not show significant spectral evolution (Fig.1). The most
straightforward interpretation for these tails is the curvature effect due to delay of
propagation of photons from large angles with respect to the line of sight (Fenimore et
al. 1996; Kumar \& Painaitescu 2000; Wu et al. 2006). In this scenario, the decay is
strictly a power law with a slope $\alpha=-(2+\beta)$ if the time zero point is set to
the beginning of the rising segment of the lightcurve (Zhang et al. 2006, see Huang et
al. 2002 for the discussion of time zero point in a different context). This model has
been successfully tested with previous data (Liang et al. 2006a).

We show here that most of the tails in our sample have significant hard-to-soft spectral
evolution (see Figs.2 and 3). The simplest curvature effect alone cannot explain this
feature. We speculate three scenarios that may result in a spectral evolution feature
and test them in turn with the data.

The first scenario is under the scheme of the curvature effect of a structure jet model.
Different from
the previous structured jet models (M\'esz\'aros, Rees \& Wijers 1998; Zhang \&
M\'{e}sz\'{a}ro 2002; Rossi et al. 2002) that invoke an angular structure of both energy
and Lorentz factor, one needs to assume that the spectral index $\beta$ is also
angle-dependent in order to explain the spectral evolution. Furthermore, in order to make
the model work, one needs to invoke a more-or-less on axis viewing geometry. Nonetheless,
this model makes a clear connection between the spectral evolution and the lightcurve, so
that $ f^{c}(\nu,t)\propto[{(t-t_p)}/{\Delta t}+1]^{-[2+\beta_c(t)]} \nu^{-\beta_c(t)}$,
where $\beta_c(t)$ is the observed spectral evolution fitting with
$\beta_c(t)=a+\kappa\log t$. We test this model with GRBs 060218 and 060614, the two
typical GRBs with strong spectral evolution, and find that it fails to reproduce the
observed light curves.

The second scenario is the superimposition of the curvature effect with a putative
underlying power-law decay emission component. This scenario is motivated by the discovery
of an afterglow-like soft component during $10^4-10^5$ seconds in the nearby GRB 060218
(Campana et al. 2006). We process the XRT data of this component, and derive a decay slope
$-1.15\pm 0.15$ and the power law photon spectral index $4.32\pm 0.18$. This soft component
cannot be interpreted within the external shock afterglow model (see also Willingale et al.
2007), and its origin is unknown. A speculation is that it
might be related to the GRB central engine (e.g. Fan et al. 2006), whose nature is a great
mystery. The most widely discussed GRB central engine is a black hole - torus system
or a millisecond magnetar. In either model, there are in principle
two emission components (e.g. Zhang \& M\'esz\'aros 2004 and references therein). One is
the ``hot'' fireball related to neutrino annihilation. This component tends to be
erratic, leading to significant internal irregularity and strong internal shocks. This
may be responsible for the erratic prompt gamma-ray emission we see. The second component
may be related to extracting the spin energy of the central black hole (e.g. Blandford \&
Znajek 1977; M\'esz\'aros \& Rees 1997b; Li 2000) or the spin energy of the central
millisecond pulsar (through magnetic dipolar radiation, e.g. Usov 1992; Dai \& Lu 1998;
Zhang \& M\'esz\'aros 2001). This gives rise to a ``cold'', probably steady Poynting flux
dominated flow. This component provides one possible reason to refresh the forward shock
to sustain a shallow decay plateau in early X-ray afterglows (Zhang et al. 2006; Nousek
et al. 2006), and it has been invoked to interpret the peculiar X-ray plateau afterglow
of GRB 070110 (Troja et al. 2007).
These fact make us suspect that at least some of the observed
spectrally evolving tails may be due to the superposition of a curvature effect tail and
an underlying soft central engine afterglow\footnote{O'Brien et al. (2006) and Willingale
et al. (2007) interpret the XRT lightcurves as the superpositions between a prompt
component and the afterglow component. The putative central engine afterglow component
discussed here is a third component that is usually undetectable but makes noticeable
contribution to the tails. }. In order to explain the observed
hard-to-soft spectral evolution the central engine afterglow component should be much
softer than the curvature effect component and it gradually dominates the observed tails.
Analogous to forward
shock afterglows, we describe the central engine afterglow component with
\begin{equation}
\label{CE} f^{u}(\nu,t)\propto t^{-\alpha_u}\nu^{-\beta_u},
\end{equation}
so that the total flux density can be modelled as
\begin{equation}\label{LC}
f(\nu,t)=f^{c}(\nu,t)+f^{u}(\nu,t),
\end{equation}
where $f^{c}(\nu,t)$ is the normal curvature effect component.
The spectral index in the XRT band at a given time thus is derived through fitting the
spectrum of $\nu f_\nu(t)$ versus $\nu$ with a power law, and the observed XRT light
curve can be modeled by
\begin{equation}\label{obs}
F_{\rm XRT}(t)=\int_{\rm XRT}[f^{c}(\nu,t)+f^{u}(\nu,t)]d\nu.
\end{equation}
We try to search for parameters to fit tails in our Group (B). Although the model can
marginally fit some of the tails, we cannot find a parameter regime to reproduce both
the lightcurves and observed spectral
index evolutions for GRBs 060218 and 060614. We therefore disfavor this model, and
suggest that the central engine afterglow emission, if any, is not significant in the
GRB tails.

The third scenario is motivated by the fact that the broad-band data of
GRB 060218 could be fitted by a cutoff power spectrum with the cutoff
energy moving from high to low energy bands (Campana et al. 2006; Liang
et al. 2006b). We suspect that our Group B tails could be of the similar
origin. As a spectral break gradually passes the XRT band, one can
detect a strong spectral evolution. We introduce an empirical model to
fit the data. The time dependent flux density could be modeled as
\begin{equation}\label{obs}
F_{\nu}(E,t)=F_{\nu,m}(t)\left[\frac{E}{E_c(t)}\right]^{-\beta}e^{-E/E_c(t)}
\end{equation}
where
\begin{equation}
F_{\nu,m}(t)=F_{\nu,m,0}\left(\frac{t-t_0}{t_0}\right)^{-\alpha_1}
\end{equation}
and
\begin{equation}
E_c(t)=E_{c,0}\left(\frac{t-t_0}{t_0}\right)^{-\alpha_2}
\end{equation}
are the temporal evolutions of the peak spectral density and the cutoff energy of the
exponential cutoff power law spectrum, respectively. In the model, $t>t_0$ is required,
and $t_0$ is taken as a free parameter. Physically it should roughly correspond to the
beginning of the internal shock emission phase, which is near the GRB trigger time. Our
fitted $t_0$ values (Table 1) are typically 10-20 seconds, usually much earlier than the
starting time of the steep decay tails, which are consistent with the theoretical
expectation.  The evolution of $E_c$ has been measured for GRB 060218 (Camapana et al.
2006; Ghisellini et al. 2006; Liang et al. 2006b; Toma et al. 2007). We first test this
model with this burst. Our fitting results are shown in Fig. 4. We find that this model
well explains the light curve and the spectral evolution of combined BAT-XRT data of GRB
060218. We therefore apply the model to both the light curves and spectral evolution
curves of other Group B tails as well (Fig.2). We do not fit Group C tails (Fig.3)
because of the flare contamination. Our fitting results\footnote{In
principle one should derive the parameters with the combined best fits to both the
light curves and $\beta$ evolutions. This approach is however impractical since the
degrees of freedom of the two fits are significantly different. We therefore fit the
light curves first, and then refine the model parameters to match the spectral
evolution behaviors . The $\chi^2$ reported in Table 1 are calculated with the refined
model parameters for the light curves. We cannot constrain the uncertainties and uniqueness of the model
parameters with this method} are displayed in Fig.2 and are tabulated in Table 1 . The
$\chi^2$ and the degrees of the freedom of the fitting to the light curves are also
marked in Fig.2. Although the flickering features in some light curves make the reduced
$\chi^2$ much larger than unity, the fittings are generally acceptable, indicating that
this model is a good candidate to interpret the data. The distributions of the fitting
parameters are shown in Fig.5. The typical $E_{c,0}$ is about 90 keV at $t_0\sim 16$
seconds. The distribution of the peak spectral density decay index $\alpha_1$ has more
scatter than the $E_c$ decay index $\alpha_2$. Interestingly it is found that $\alpha_1$
is strongly correlated with $\alpha$, say, $\alpha_1=(0.82\pm 0.10)\alpha-(1.00\pm
0.38)$ (see Fig. 6; the quoted errors are at $1\sigma$ confidence level.), with a
Spearman correlation coefficient $r=0.90$ and a chance probability $p<10^{-4}$ ($N=16$).
This is the simple manifestation of the effect that the faster a burst cool (with a
steeper $\alpha_1$), the more rapidly the tail drops (with a steeper $\alpha$). The
$\alpha_2$ parameter is around $1.4$ as small scatter. This indicates that the evolution
behaviors of $E_c$ are similar among bursts, and may suggest a common cooling process
among different bursts.

Comparing the three scenarios discussed above, the third
empirical model of the prompt emission region is the best candidate to interpret the spectral evolution of the Group B tails. The Group C tails may include additional (but weaker) heating processes during the decay
phase (Fan \& Wei 2005; Zhang et al. 2006), as have been suggested by the fluctuations
and flares on the decaying tails. The steep decay component has been also interpreted as
cooling of a hot cocoon around the jet (Pe'er et al. 2006). This model may be
relevant to some tails of the long GRBs, but does not apply to the tails from the bursts
of compact star merger origin (such as GRB 050724 and probably also GRB 060614, Zhang et
al. 2007b). Another scenario to interpret the tails is a highly radiative blast wave that
discharges the hadronic energy in the form of ultra-high energy cosmic ray neutrals and
escaping cosmic-ray ions (Dermer 2007). It is unclear, however, whether the model can
simultaneously interpret both the observed lightcurves and the spectral evolution curves
of these tails. In addition, dust scattering may explain some features of the tails,
including the spectral evolution, for some bursts (Shao \& Dai 2007).

Recently, Butler \& Kocevski (2007) used the evolution of the hardness ratio as an indicator to discriminate the GRB tail emission and the forward shock emission. As shown
in Fig.2, the spectra of the tails are significantly different from those
of the shallow decay component. Spectral behaviors, including evolution of the hardness
ratio, are indeed a good indicator to separate the two emission components. However, no
significant difference was observed between the spectra of the tails and the following
shallow decay component for the Group A bursts that show no significant spectral
evolution (Fig.1).

With the observation by CGRO/BATSE, it was found that the prompt GRBs tend to show a
spectral softening and a rapid decay (Giblin et al. 2002; Connaughton 2002). Ryde \&
Svensson (2002) found that about half of the GRB pulses for the BATSE data decay
approximately as $t^{-3}$, and their $E_p$'s also decay as a power law. These results
are consistent with the study of X-ray tails in this paper, suggesting a possible
common origin of the spectral evolution of GRB emission.

\section{CONCLUSIONS}
We have systematically analyzed the time-dependent spectra of the bright GRB tails observed
with {\em Swift}/XRT between Feb. 2005 and Jan. 2007. We select a sample of 44 bursts.
Eleven tails (Group A) in our sample are smooth and without superimposing significant flares,
and their spectra have no significant evolution features. We suggest that these tails are
dominated by the curvature effect of the prompt gamma-rays. More interestingly,
33 tails in our sample show clear hard-to-soft spectral evolution, with 16 of them (Group B)
being smooth tails directly following the prompt GRBs while the other 17 (Group C)
being superimposed with significant flares. We focus on the Group B tails
and consider three toy models to interpret the spectral evolution effect.
We find that the curvature effect of a structured jet and the superposition model with
the curvature effect and a putative underlying soft emission component cannot interpret
all the data, in particular the strong evolution observed in GRB 060218 and GRB 060614. A third empirical model invoking an apparent evolution of a cutoff
power law spectrum seems to be able to fit both the light curves and the
spectral evolution curves of the Group B tails. More detailed physical
models are called for to understand this apparent evolution effect.

\acknowledgments We acknowledge the use of the public data from the Swift data archive.
We thank an anonymous referee for helpful suggestions, and appreciate helpful discussions
with Dai Z. G., Wang X. Y., Fan Y. Z., and Qin Y. P. This work is supported by NASA
through grants NNG06GH62G, NNG06GB67G, NNX07AJ64G and NNX07AJ66G (for B.B.Z., E.W.L,, B.Z.), and by
the National Natural Science Foundation of China (for E.W.L.; grant 10463001).

 \begin{table}[]
  \caption[]{Fitting Results with the empirical model  for Group B GRB Tails}
  \label{Tab:1}
  \begin{center}\begin{tabular}{lllllllll}
  \hline\noalign{\smallskip}
GRB & $\alpha$ &$E_{c,0}  $ \  (keV) & $\beta$ & $\alpha_{1} $  & $\alpha_{2} $ & $t_{0}\ (s) $& ${\chi}^2$&$dof$ \\
  \hline\noalign{\smallskip}
050421&  2.8& 70.2& -0.8&  1.3&  1.4& 14.8&                85.3&          49\\
050724&  2.2& 83.8&  0.3&  0.4&  1.6& 25.8&               221.6&         113\\
050814&  3.2&113.6&  0.5&  1.6&  1.3& 18.4&               108.2&          72\\
050915B&  5.3& 89.3&  1.2&  4.1&  1.3& 17.6&                88.9&          75\\
051227&  2.5& 62.0&  0.4&  1.1&  1.5& 36.9&                32.6&          19\\
060115&  3.2& 81.1&  0.3&  1.7&  1.2& 16.4&                76.9&          35\\
060211A&  4.2& 81.1&  0.4&  2.6&  1.2& 22.5&                83.0&          48\\
060218&  2.2&113.9&  0.2&  1.1&  1.0& 75.5&               273.0&         211\\
060427&  3.5& 83.8&  0.8&  2.0&  1.4& 16.2&                55.9&          36\\
060428B&  4.7& 78.4&  1.2&  2.3&  1.4& 34.1&                85.0&          56\\
060614&  3.3&127.6&  0.0&  1.8&  1.4& 17.6&               871.0&         618\\
060708&  4.2& 72.9&  1.6&  2.2&  1.0&  7.3&                19.7&          19\\
061028&  4.6& 75.6& 0.0&  3.4&  1.0& 25.0&                49.9&          30\\
061110A&  4.8& 83.8&  0.7&  2.4&  1.2&  6.8&                52.5&          48\\
061222A&  4.7& 64.7&  1.2&  2.3&  1.3& 22.5&                57.3&          45\\
070110&  2.4&146.4&  0.5&  1.0&  1.2& 11.5&                77.6&          63\\
  \noalign{\smallskip}\hline
  \end{tabular}\end{center}
\end{table}

\clearpage

\renewcommand\thefigure{1}
\begin{figure}
\includegraphics[angle=0,scale=.32]{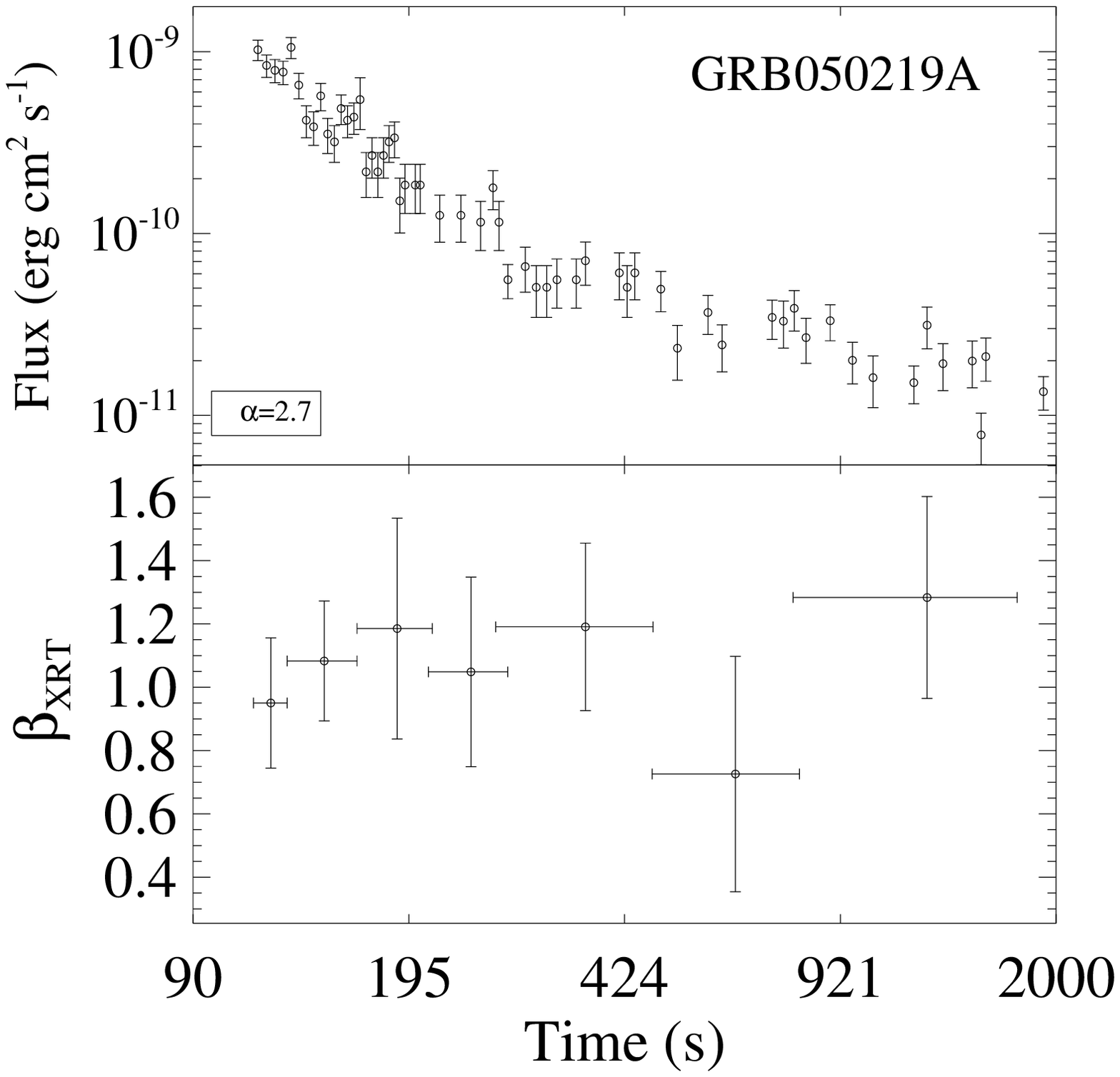}
\includegraphics[angle=0,scale=.32]{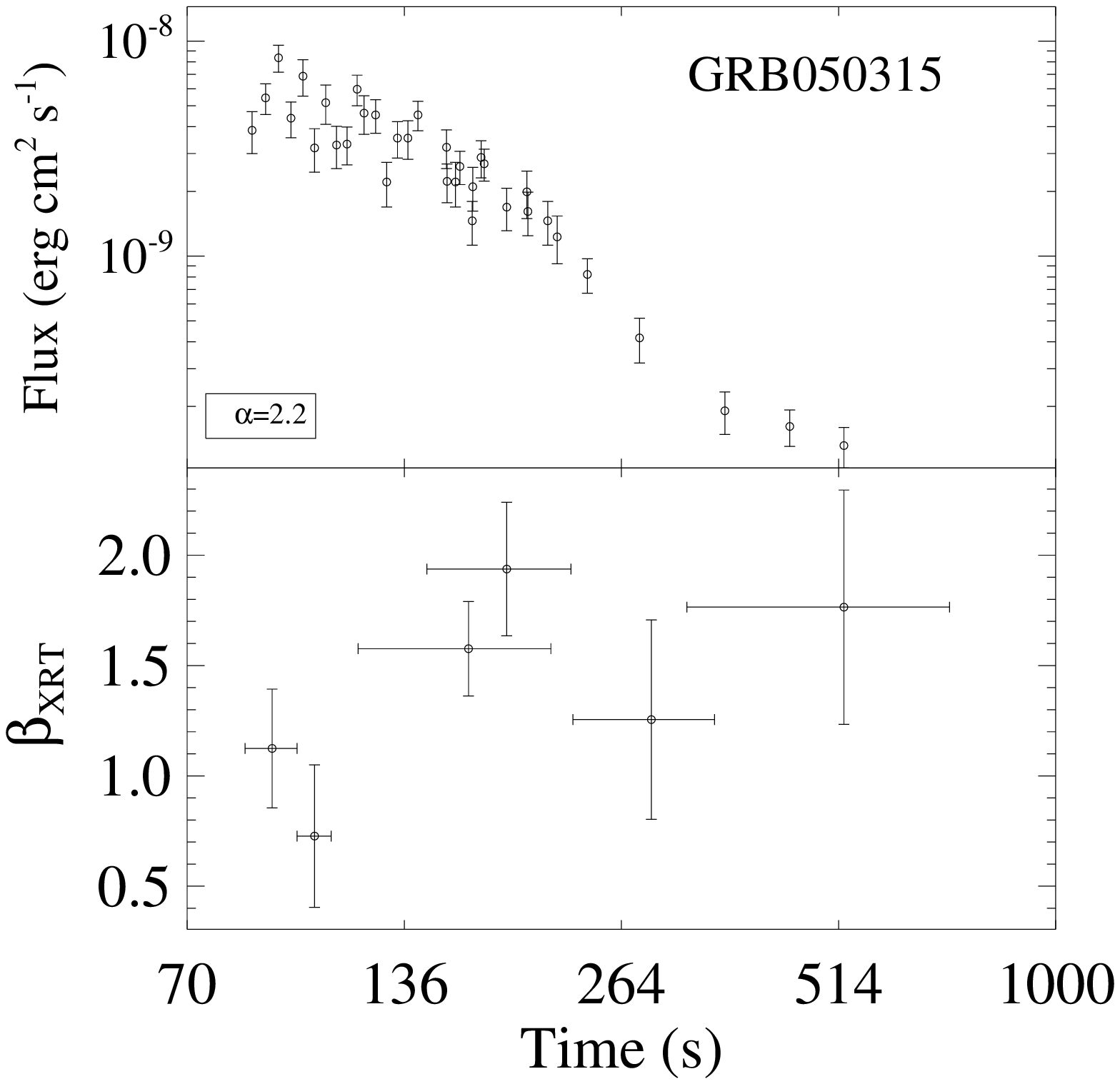}
\includegraphics[angle=0,scale=.32]{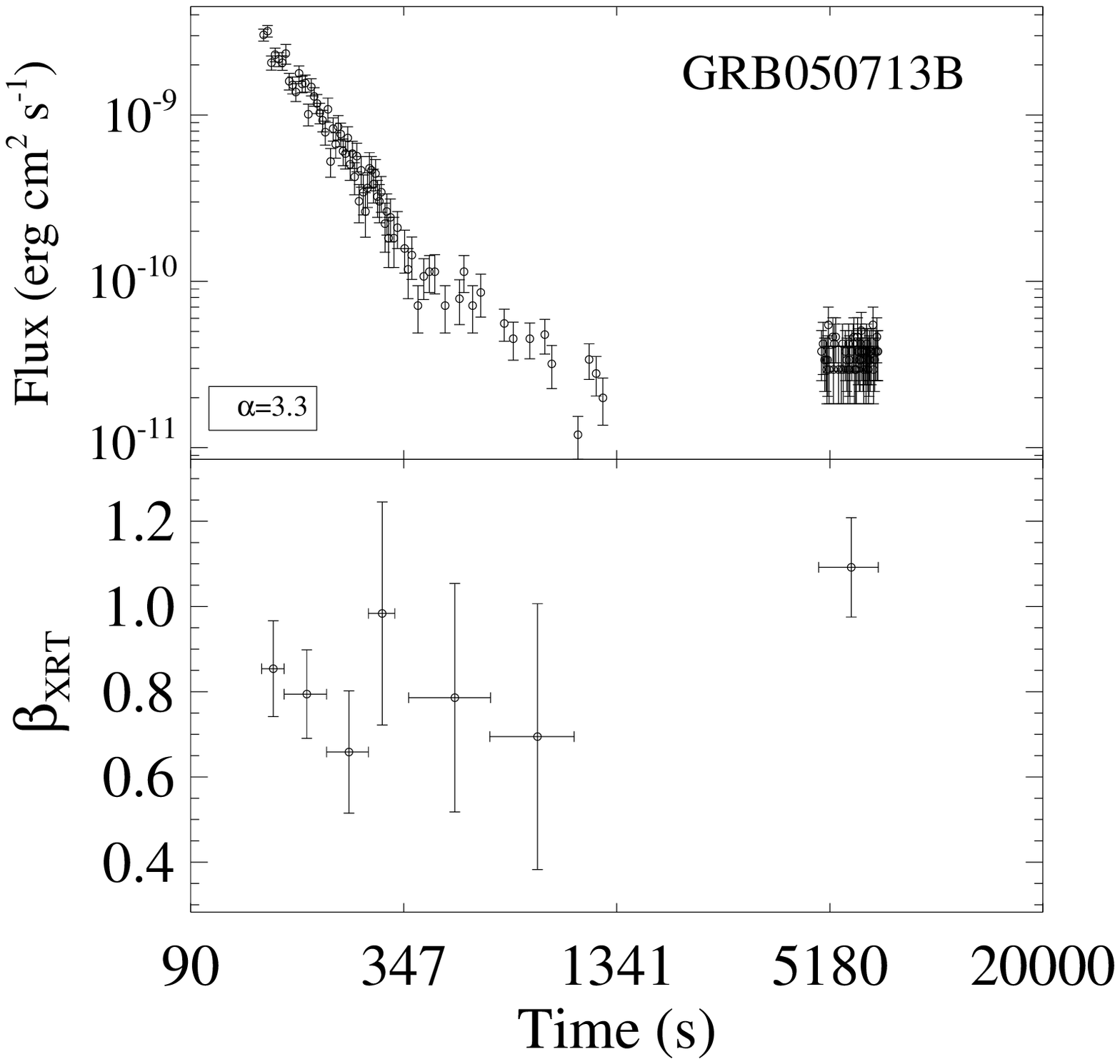}
  \hfill
\includegraphics[angle=0,scale=.32]{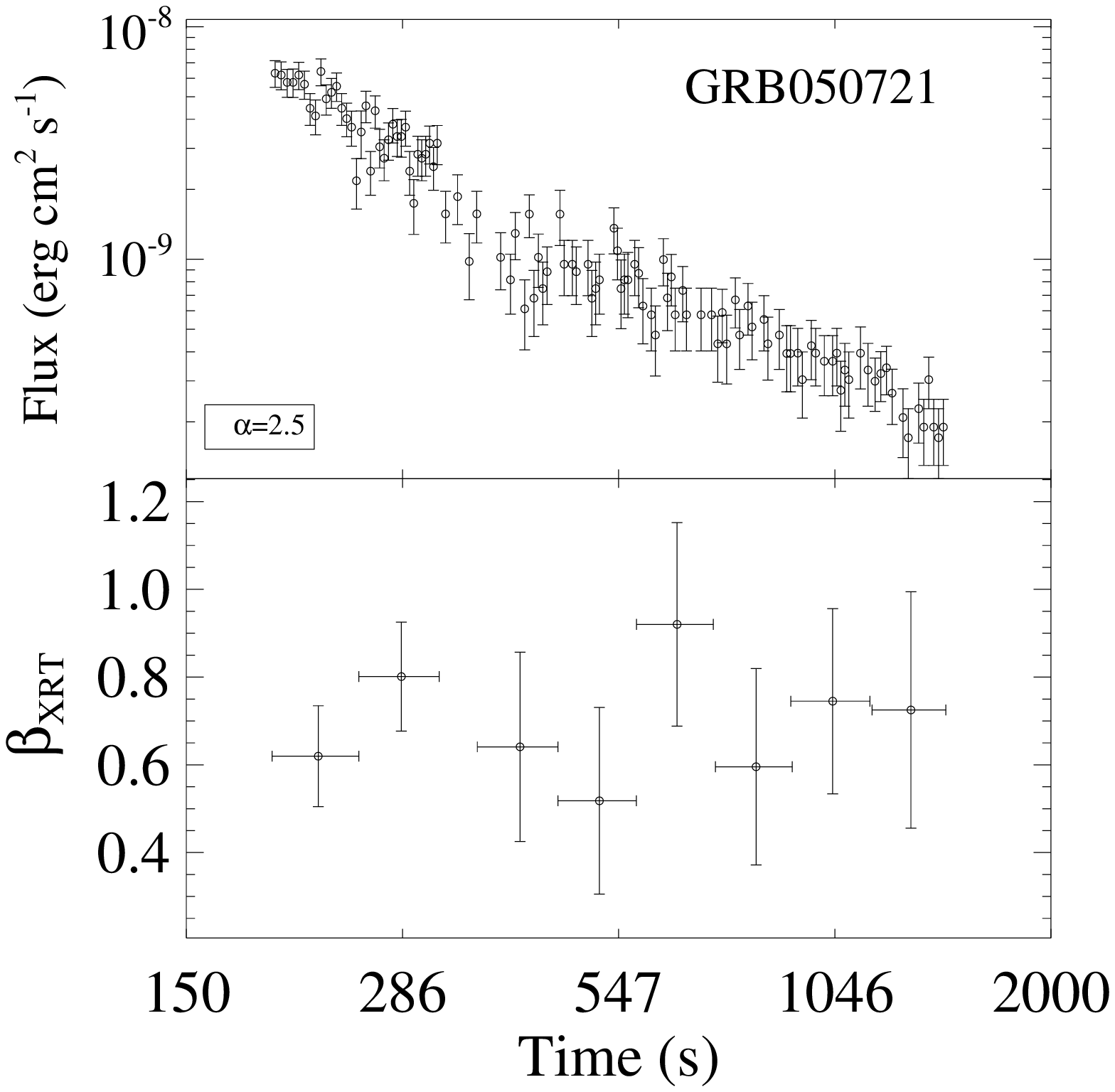}
\includegraphics[angle=0,scale=.32]{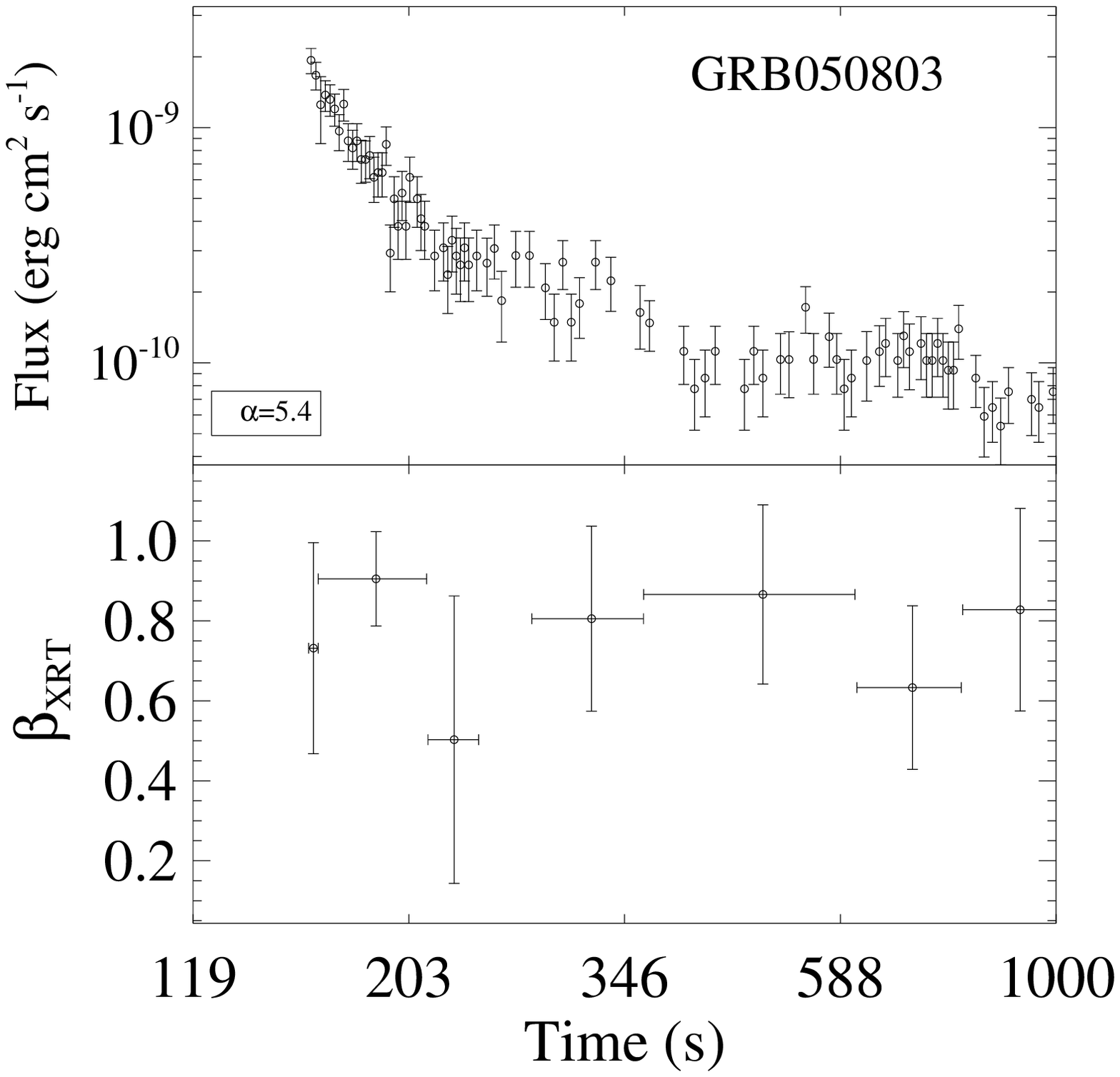}
\includegraphics[angle=0,scale=.32]{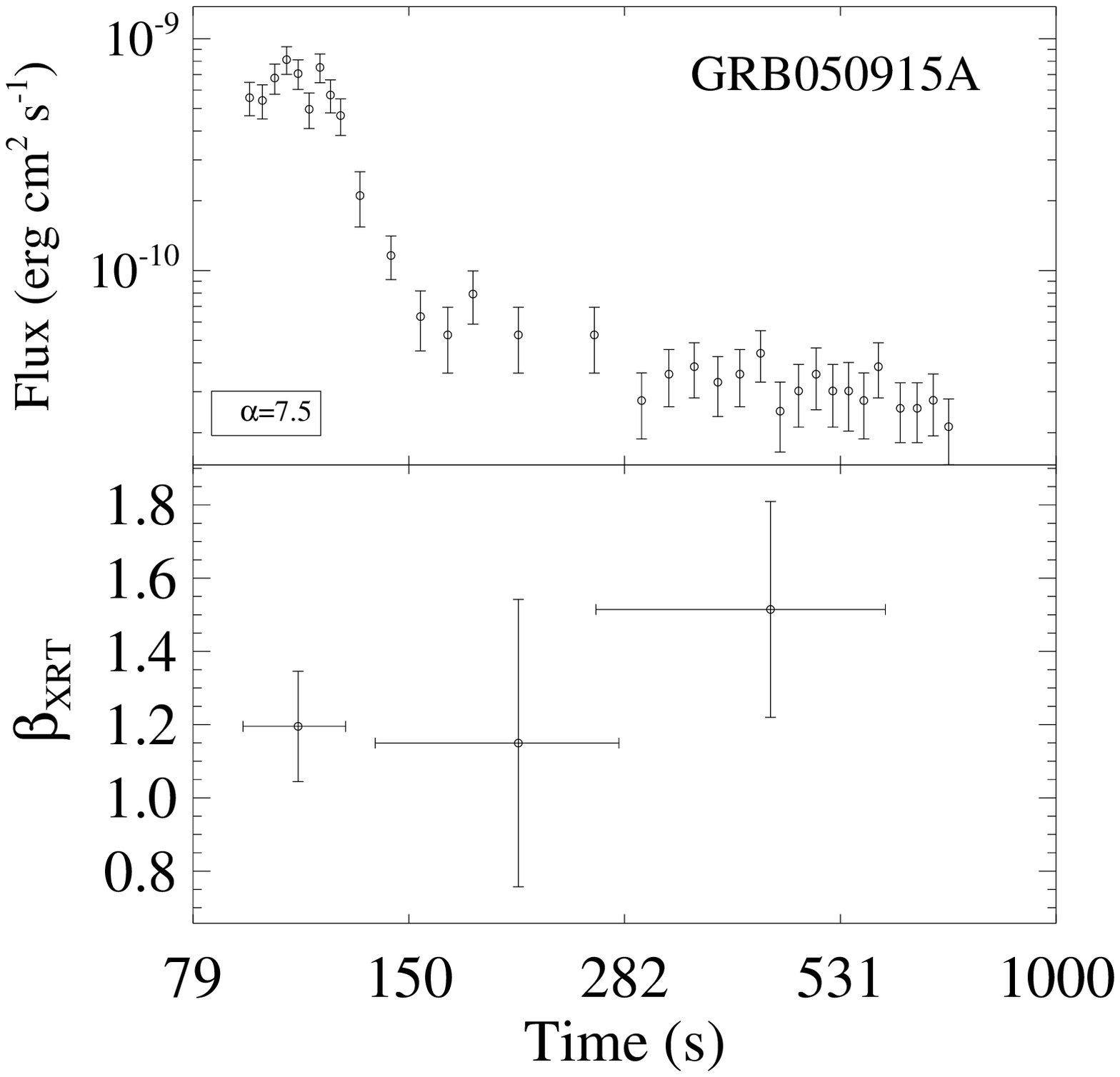}
\includegraphics[angle=0,scale=.32]{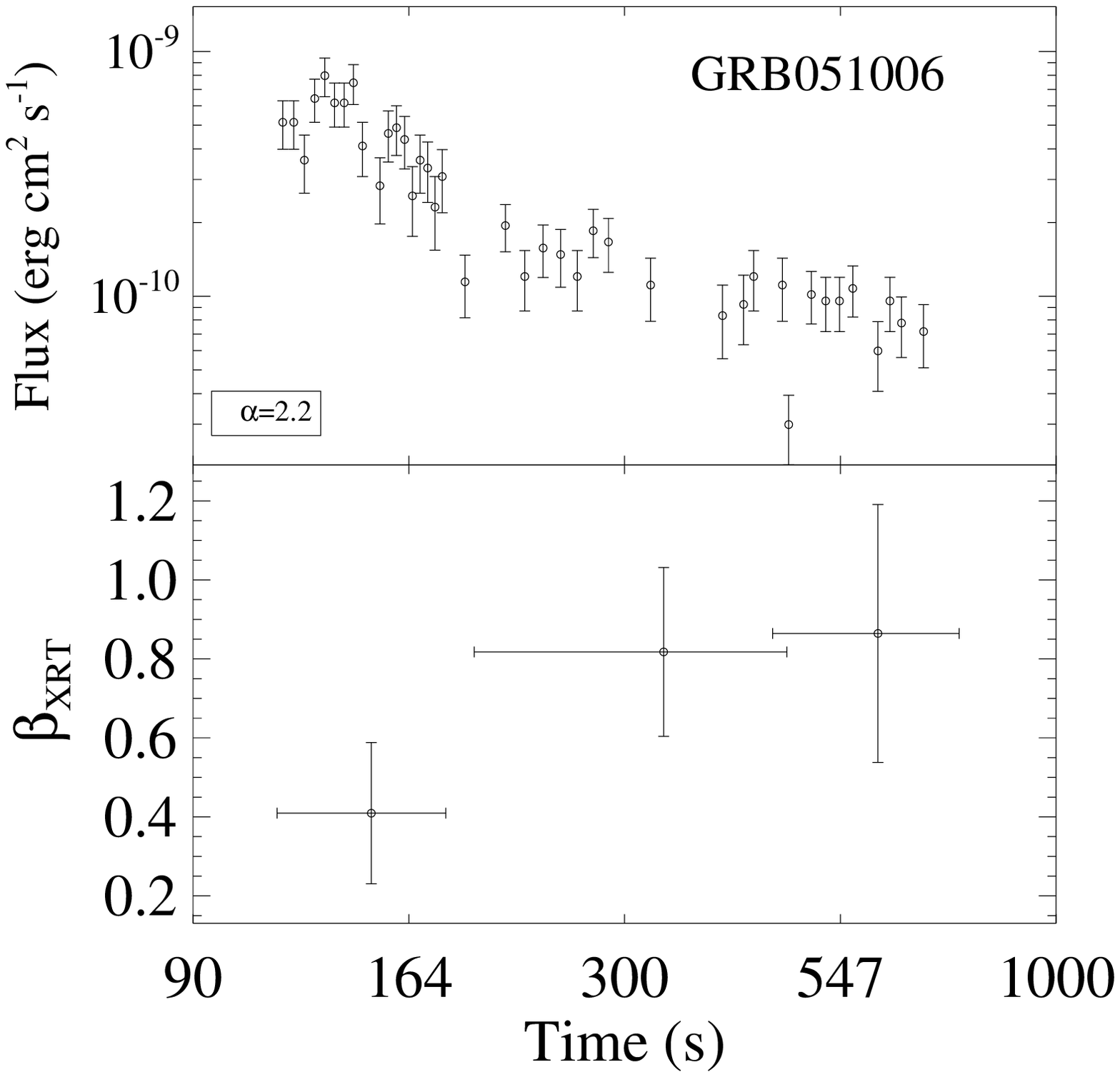}
 \hfill
\includegraphics[angle=0,scale=.32]{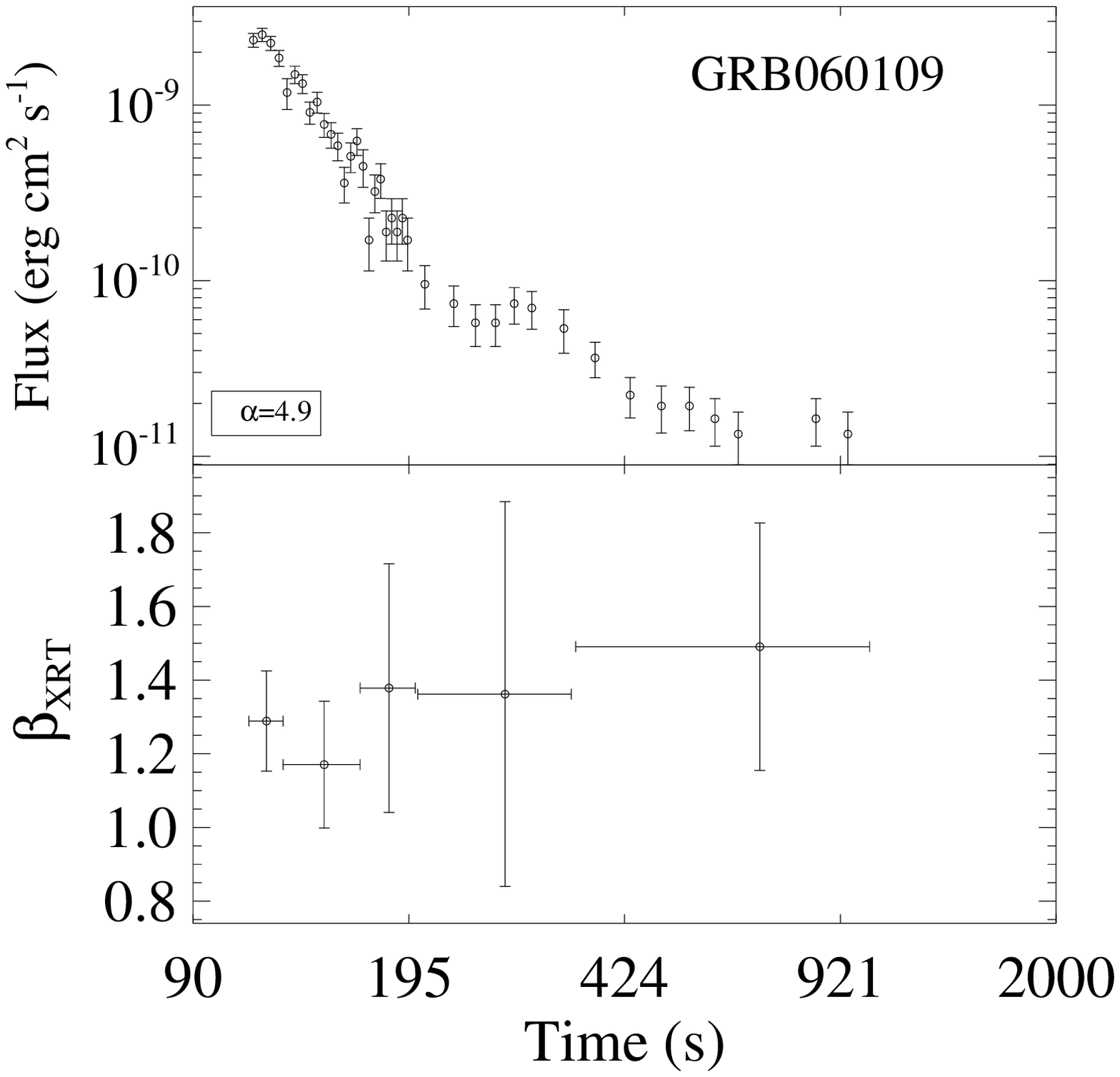}
\hfill
\includegraphics[angle=0,scale=.32]{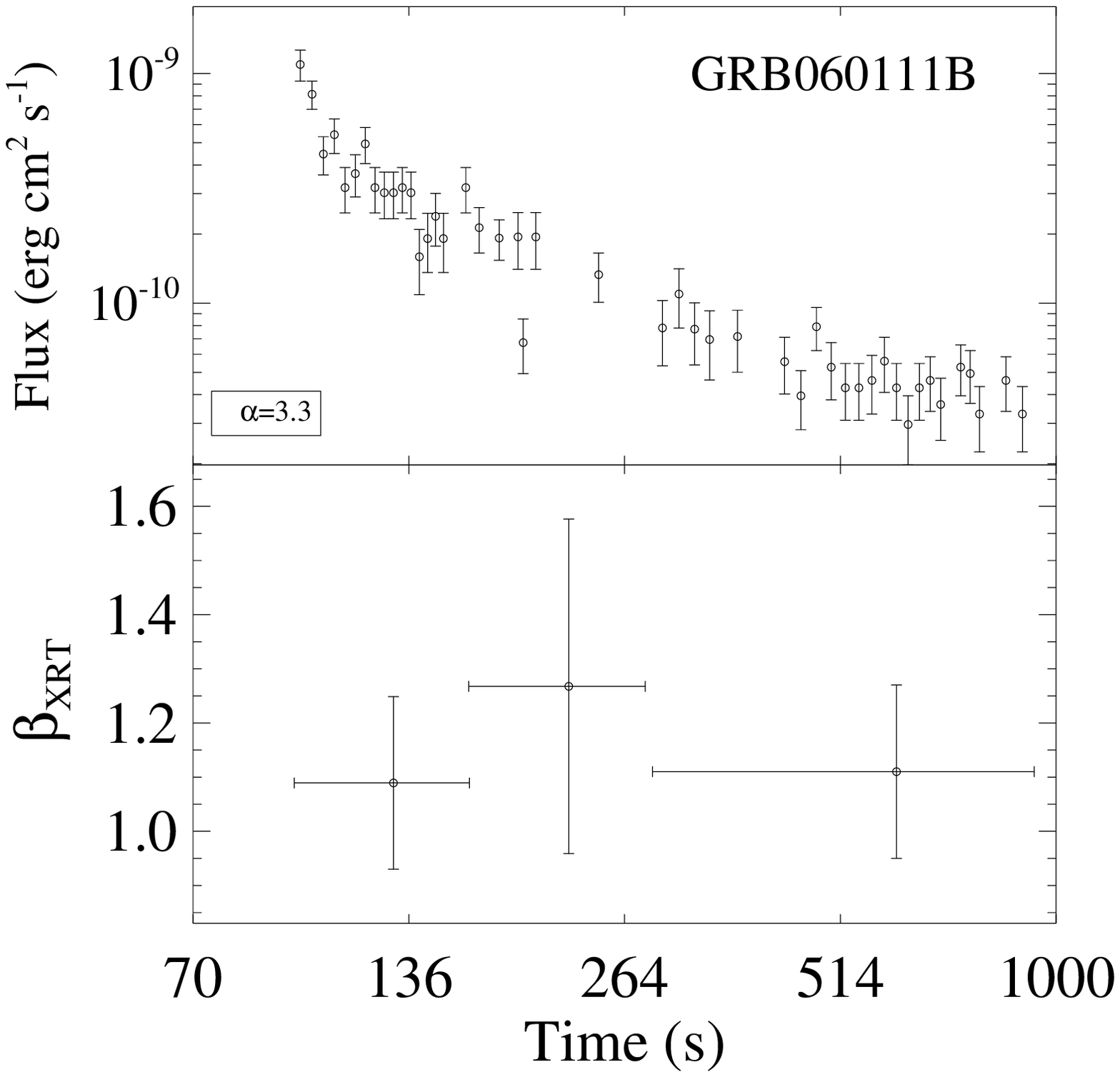}

\includegraphics[angle=0,scale=.32]{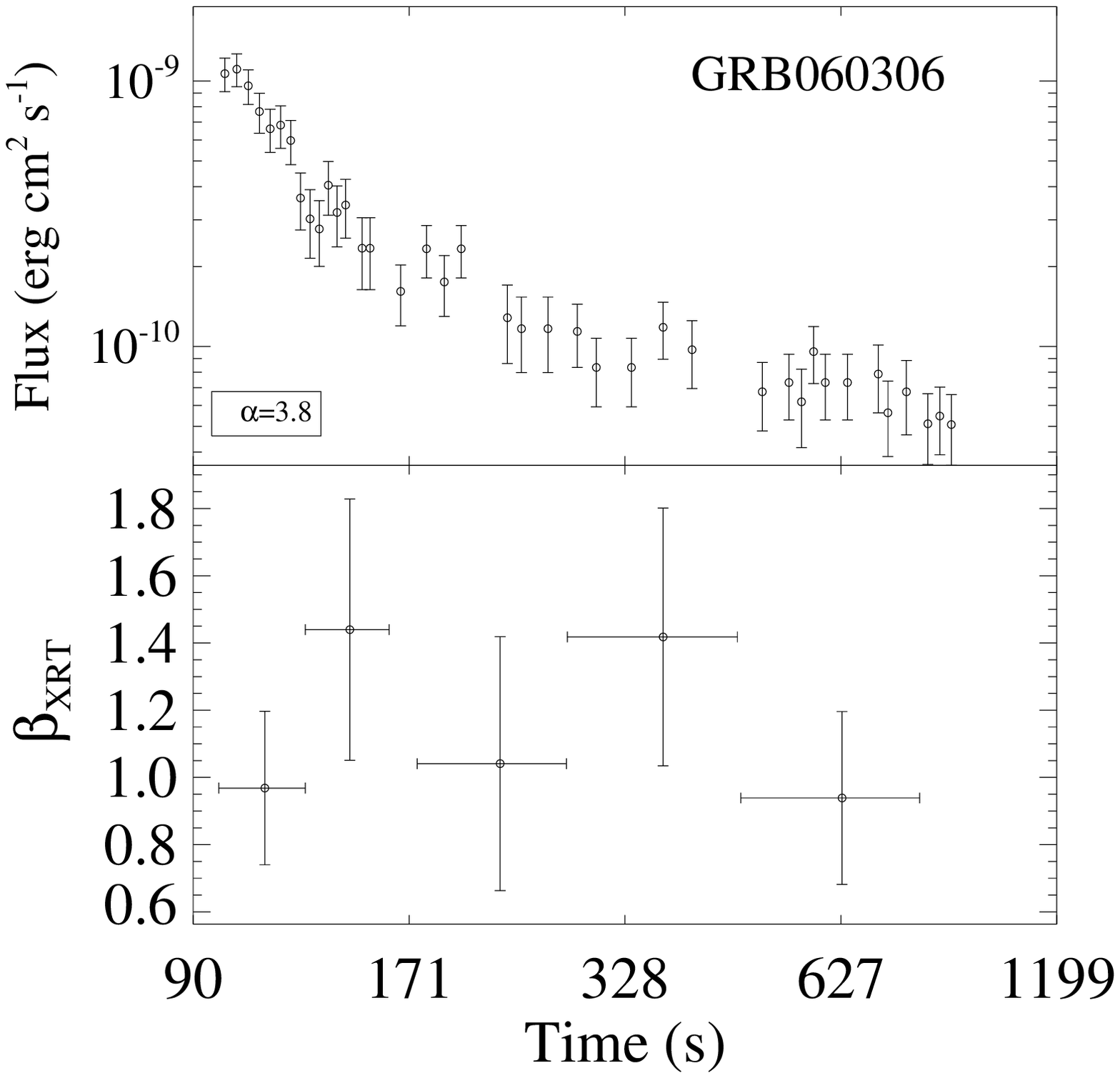}
  \hfill
\includegraphics[angle=0,scale=.32]{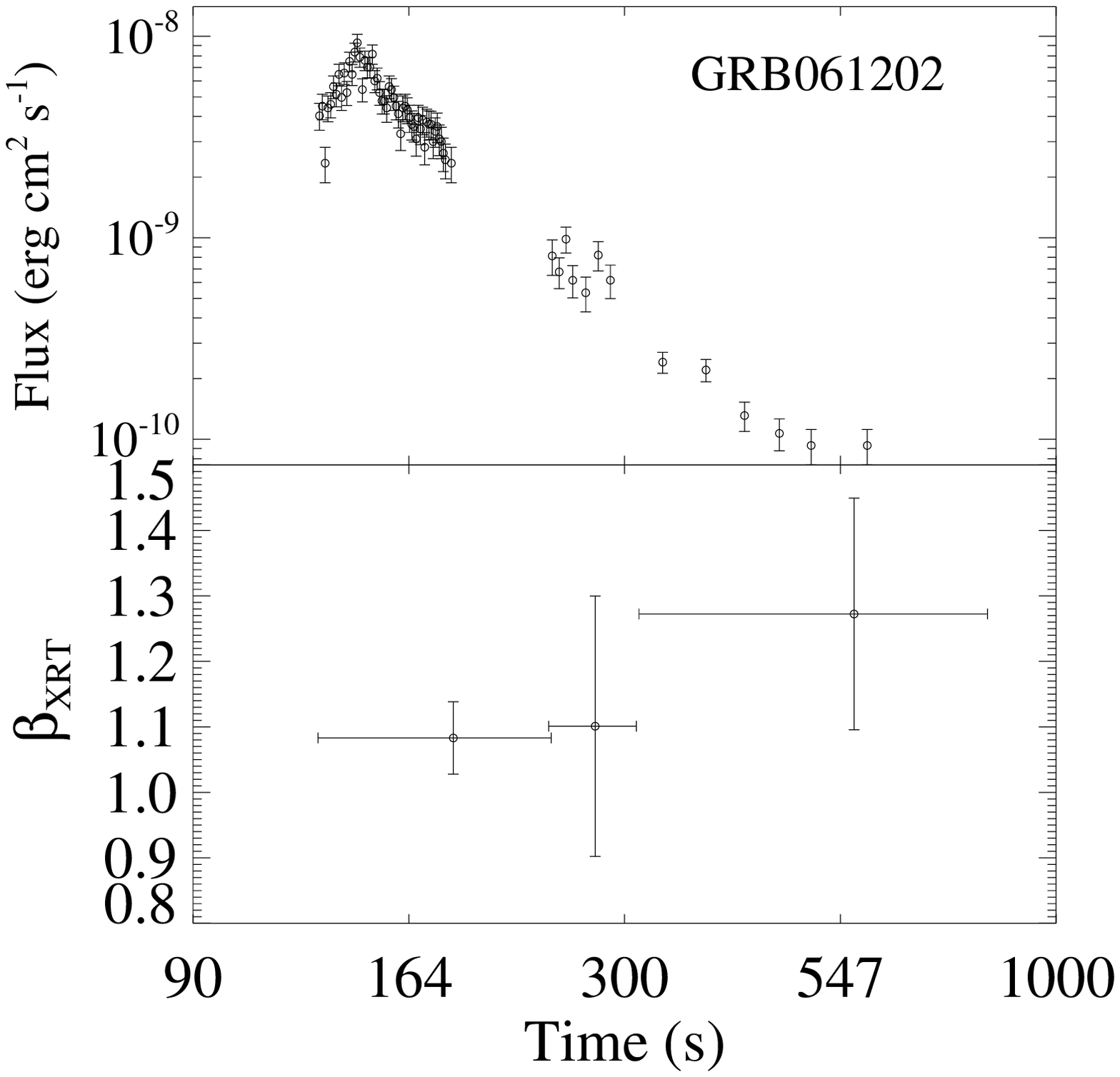}
\hfill
\hfill
\hfill
\hfill
\hfill
\hfill
\hfill
\hfill
\caption{XRT light curve (upper panel of each plot) and spectral index as a function of
time (lower panel of each plot) for those tails without significant spectral evolution
(Group A). The horizontal error bars in the lower panels mark the time interval for the
spectral analyses. Whenever available, the shallow decay segments following the tails and
their spectral indices are also shown. }
\end{figure}

\clearpage

\renewcommand\thefigure{2}
\begin{figure}
\includegraphics[angle=0,scale=.32]{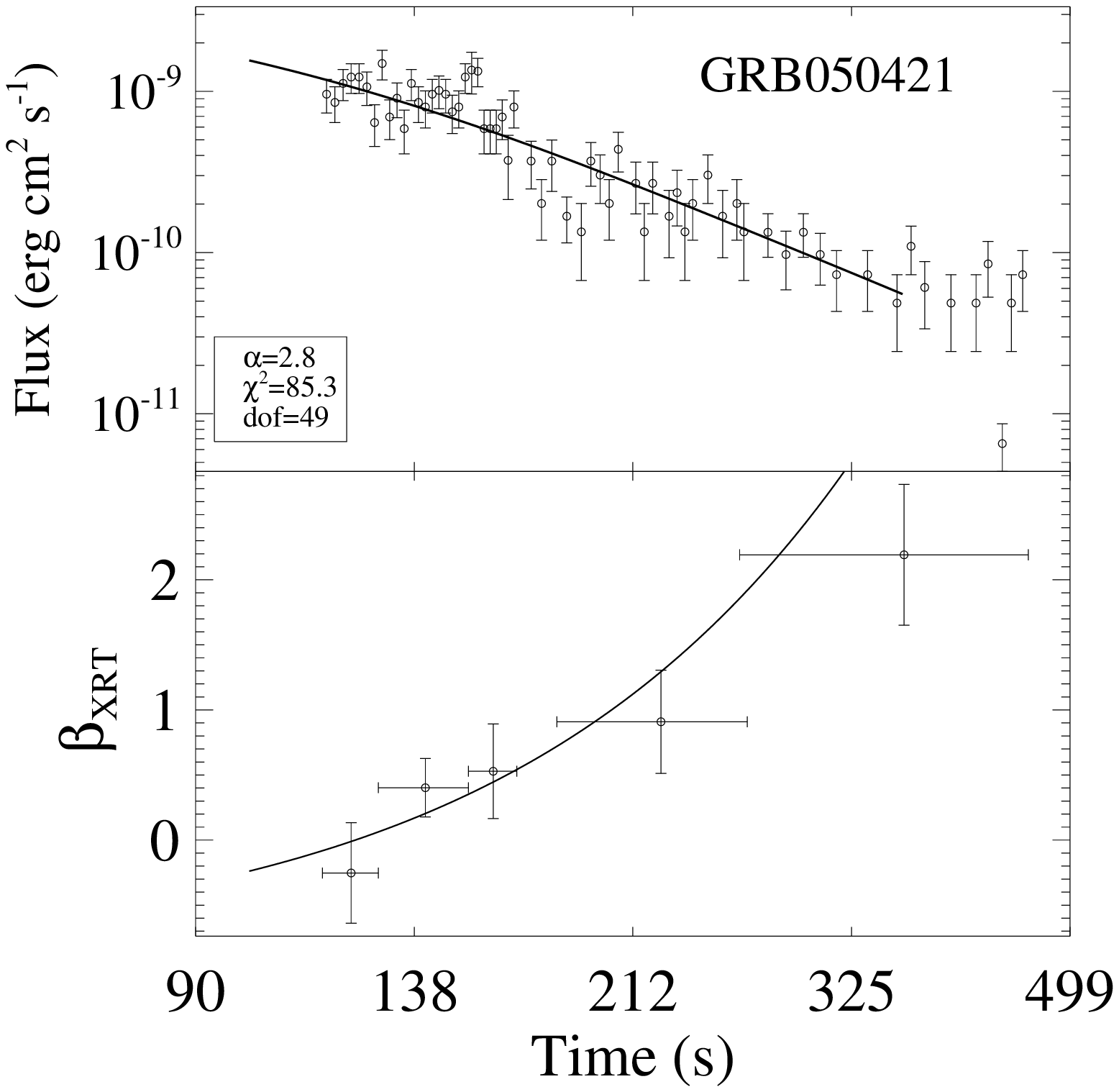}
\includegraphics[angle=0,scale=.32]{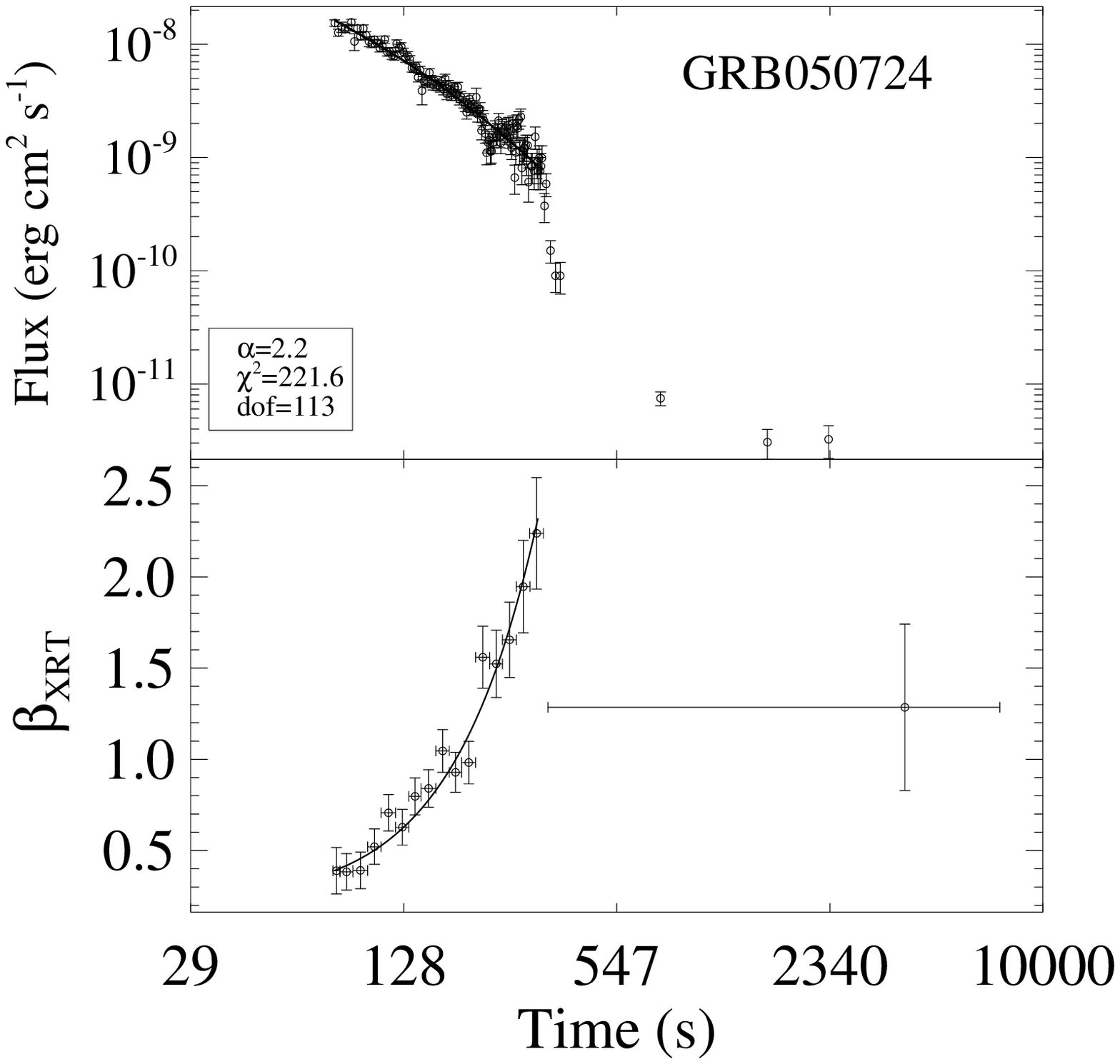}
\includegraphics[angle=0,scale=.32]{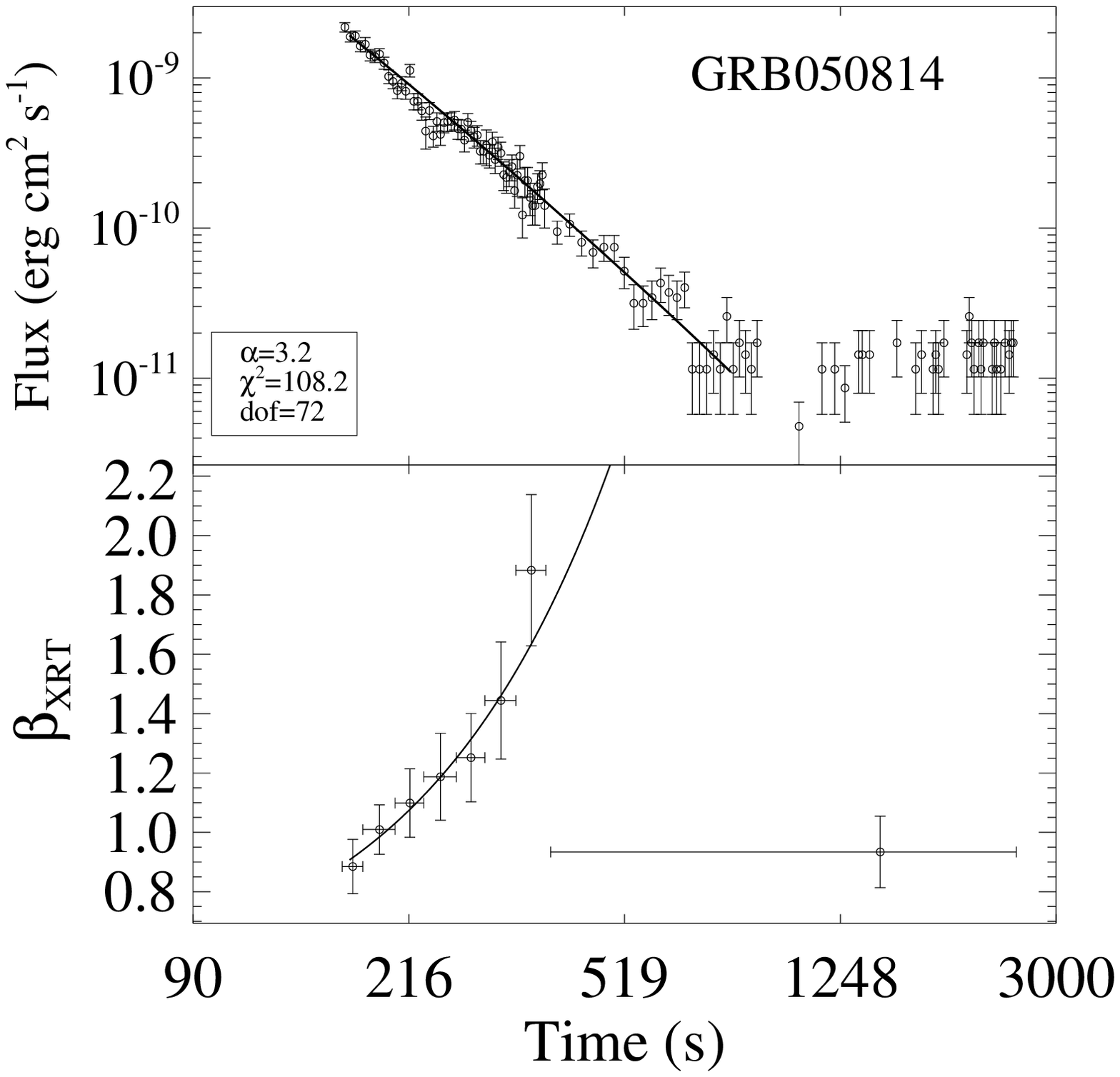}
 \hfill
\includegraphics[angle=0,scale=.32]{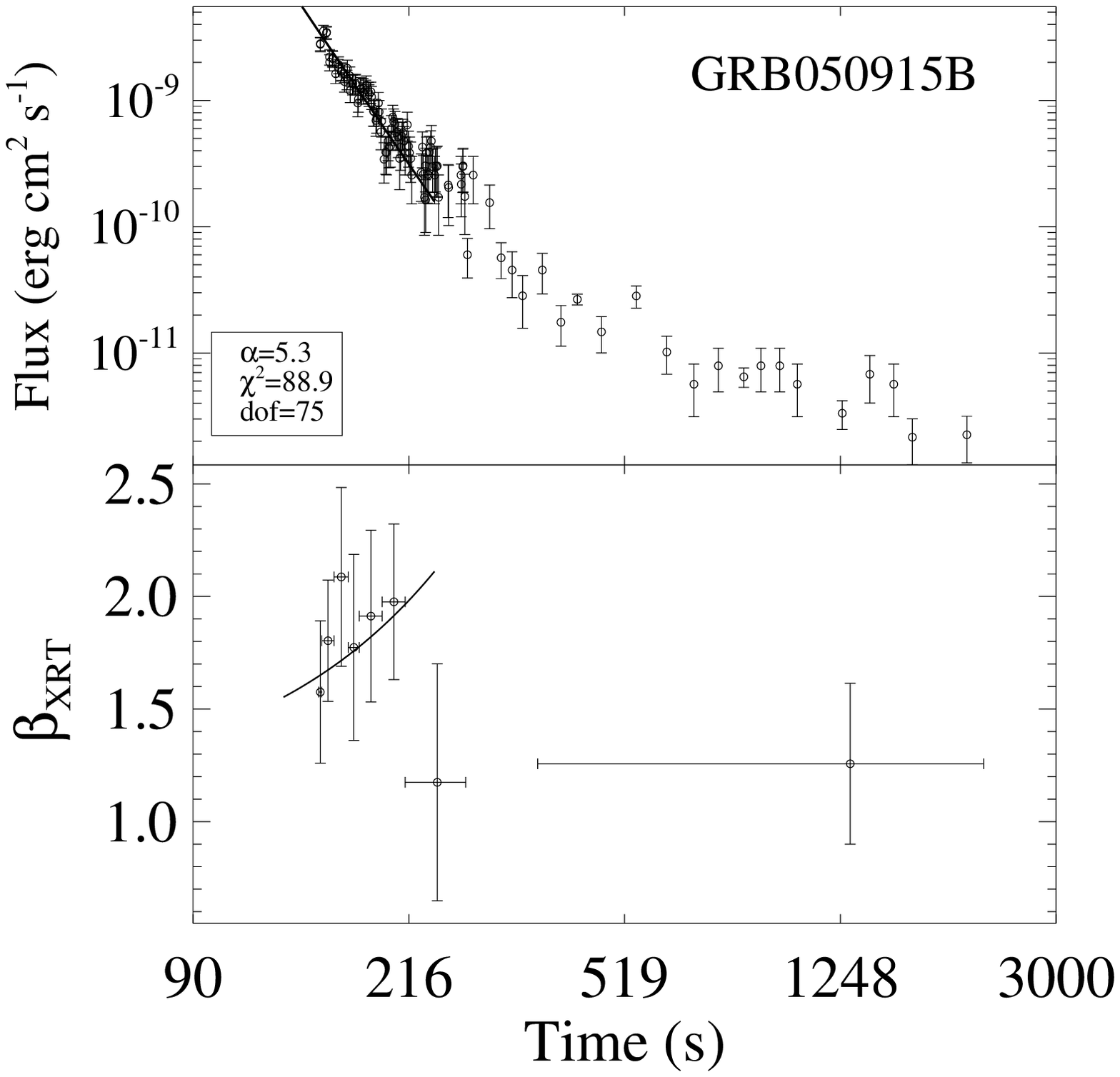}
\includegraphics[angle=0,scale=.32]{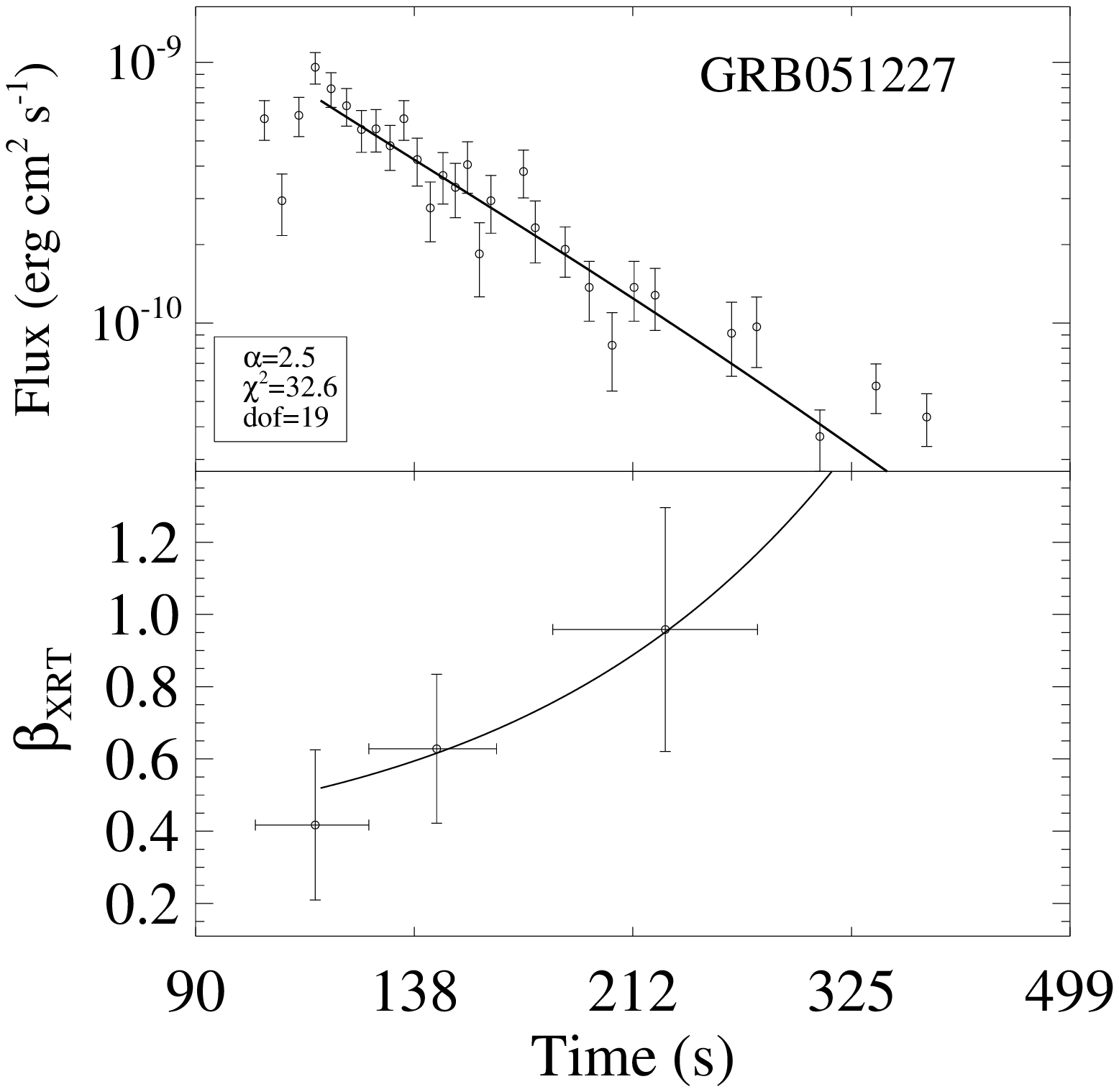}
\includegraphics[angle=0,scale=.32]{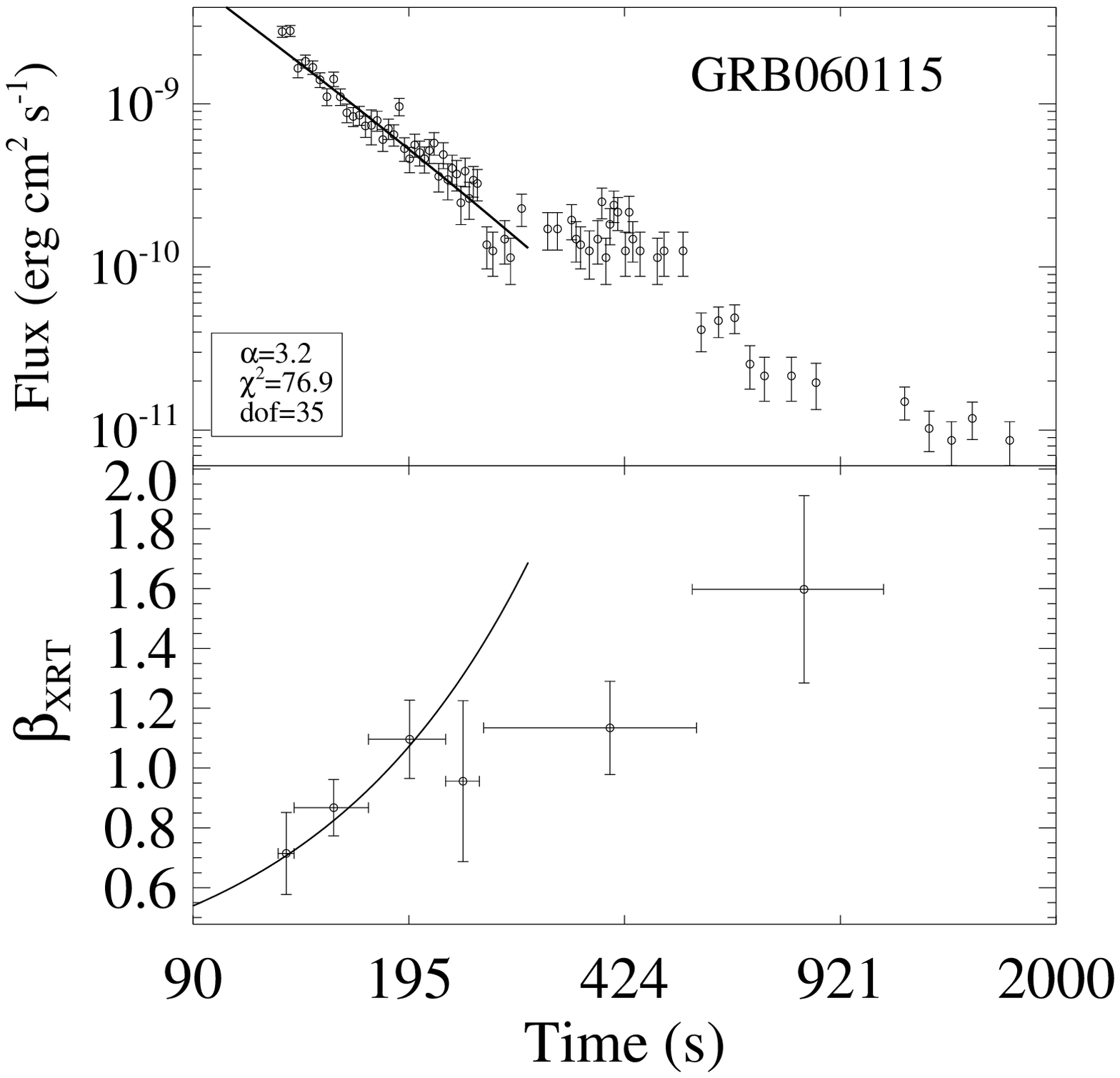}
  \hfill
\includegraphics[angle=0,scale=.32]{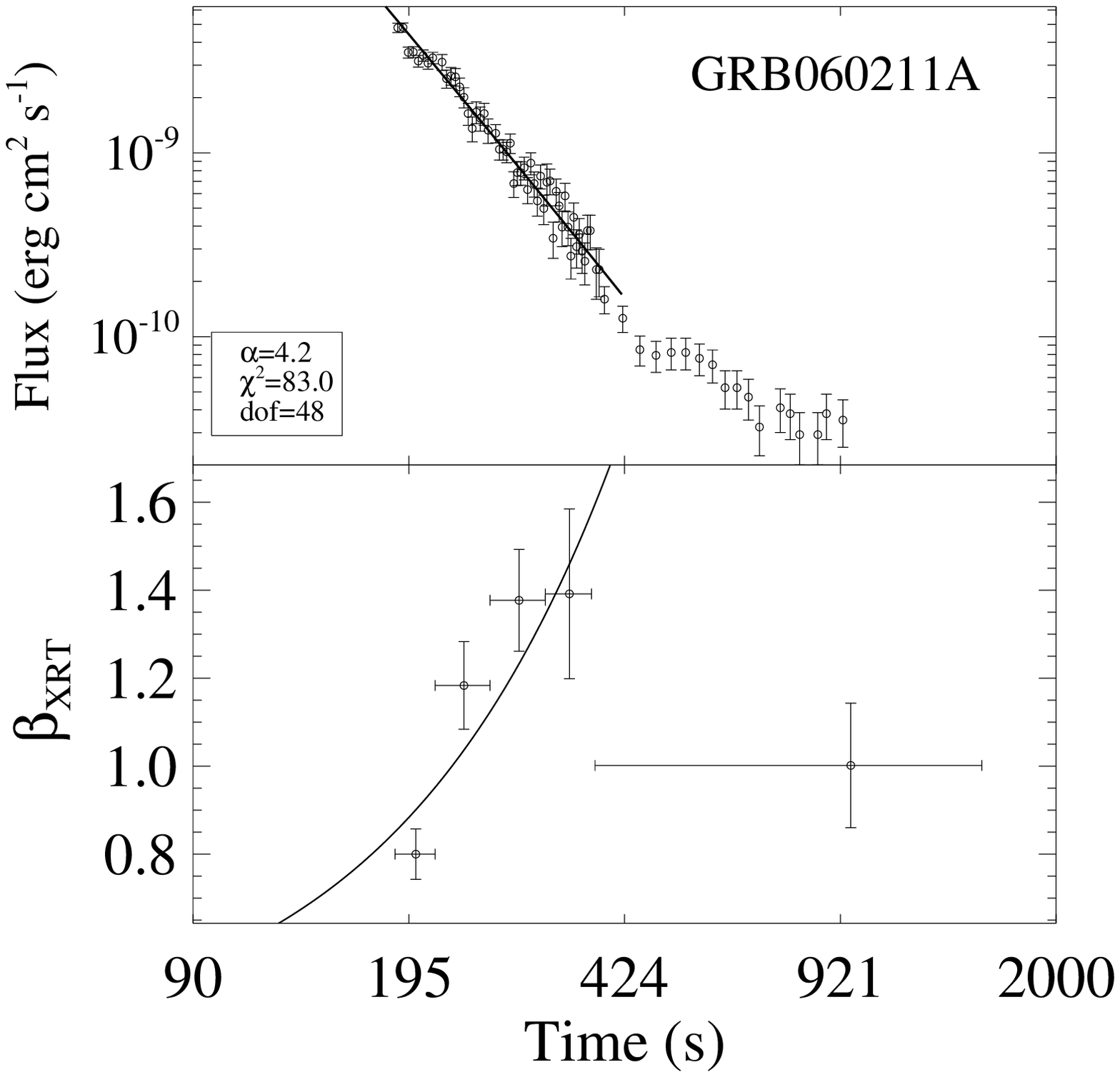}
 \includegraphics[angle=0,scale=.32]{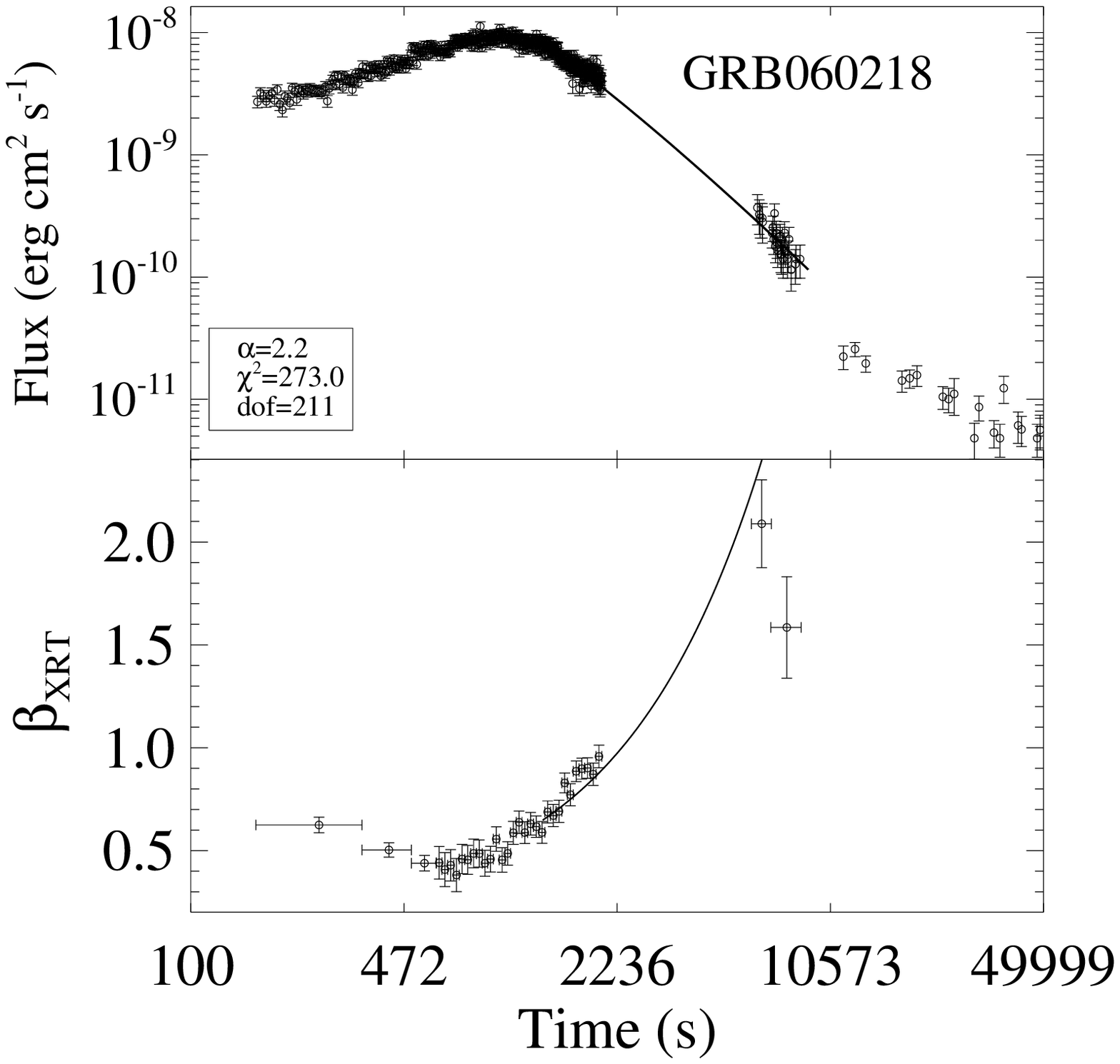}
\includegraphics[angle=0,scale=.32]{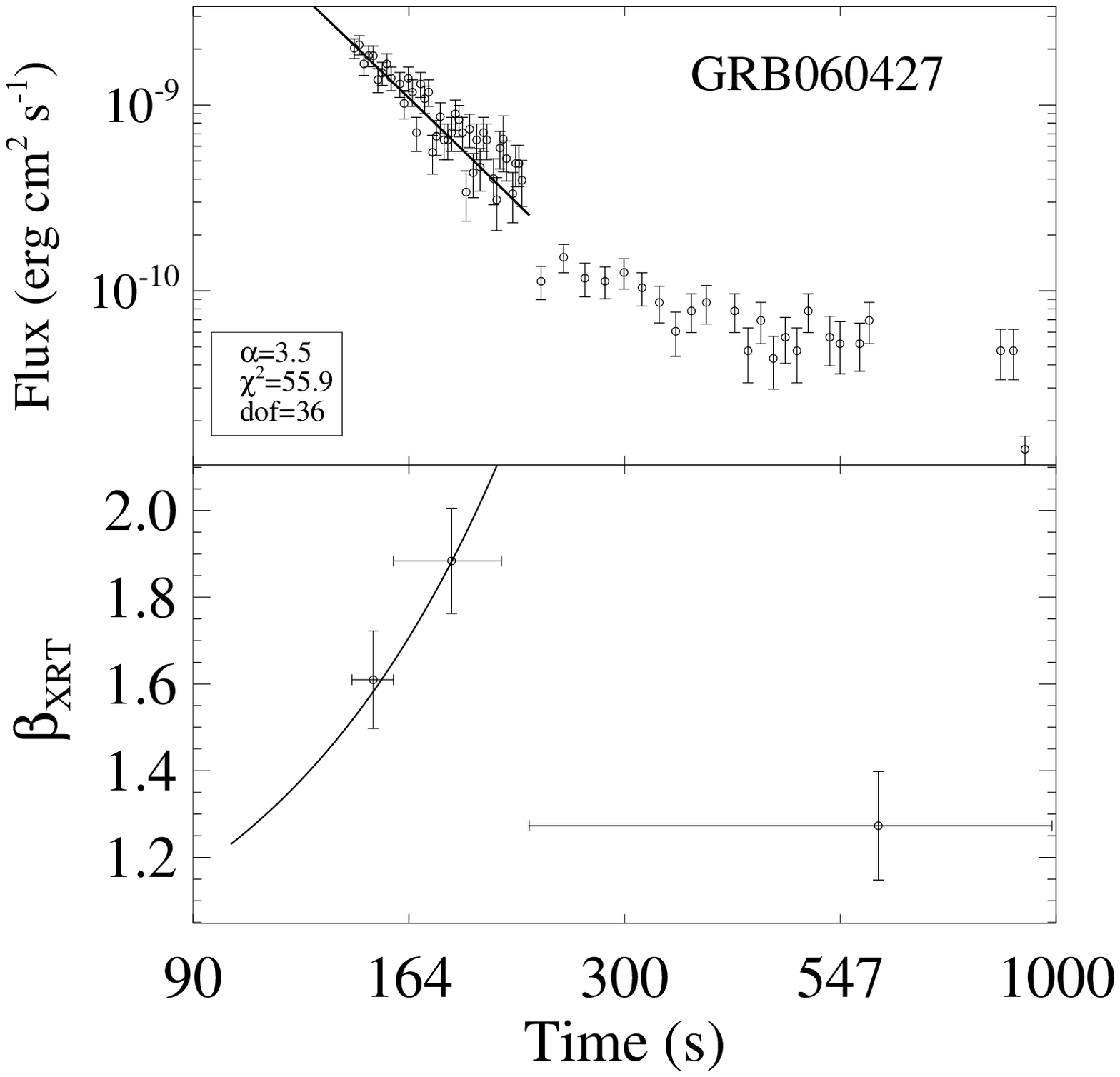}
\hfill
\includegraphics[angle=0,scale=.32]{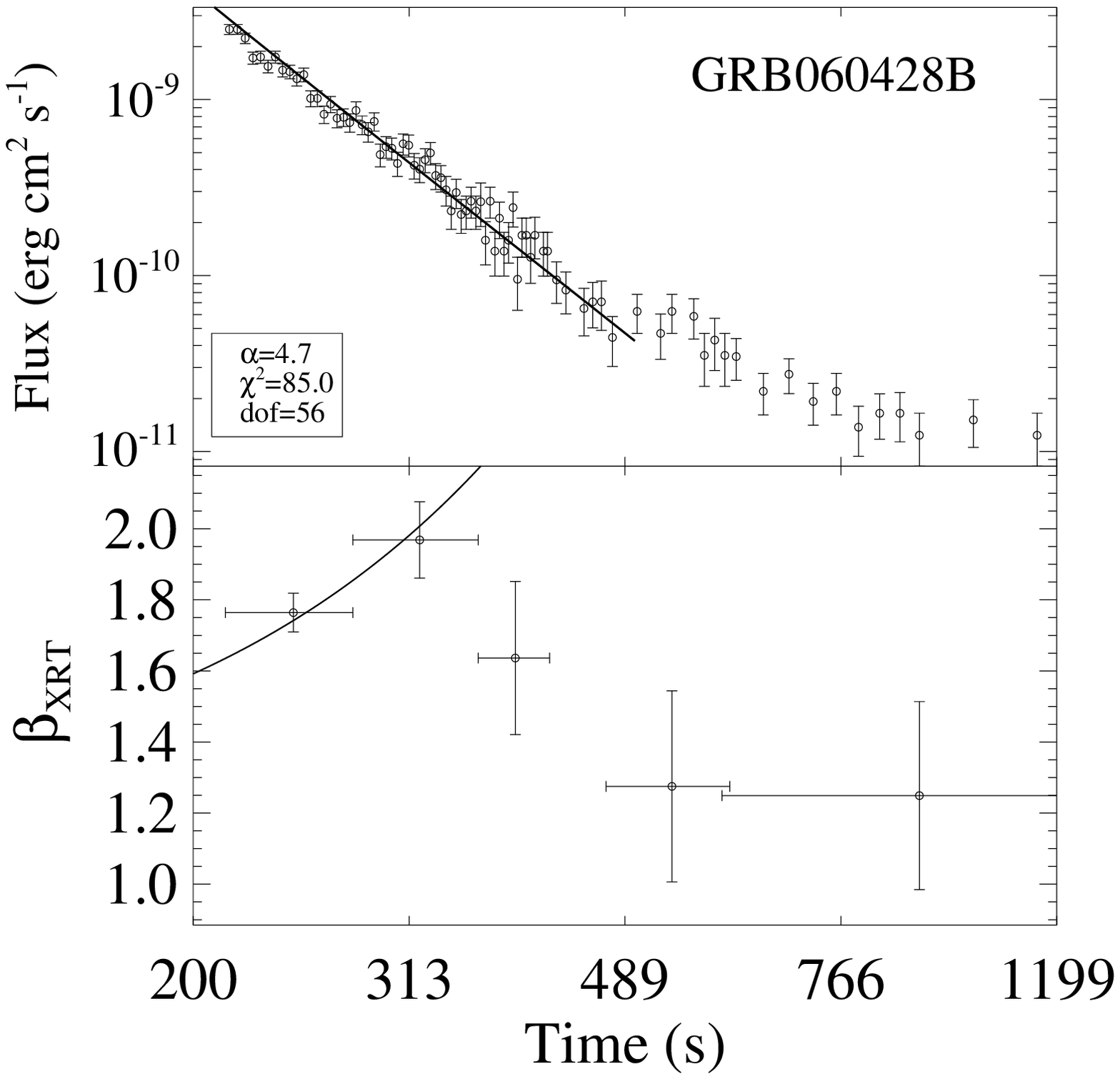}
\hfill
\includegraphics[angle=0,scale=.32]{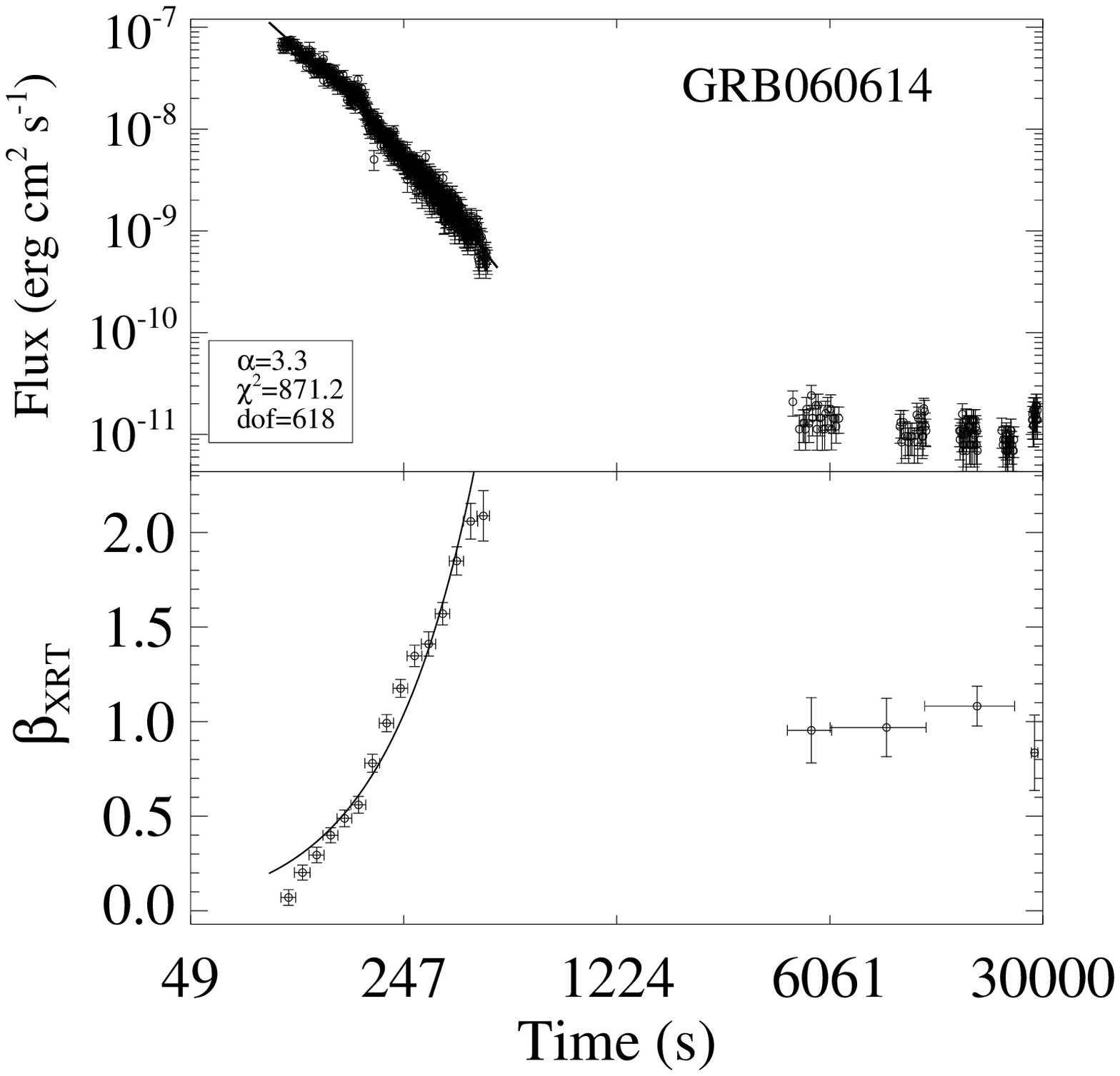}
\hfill
\includegraphics[angle=0,scale=.32]{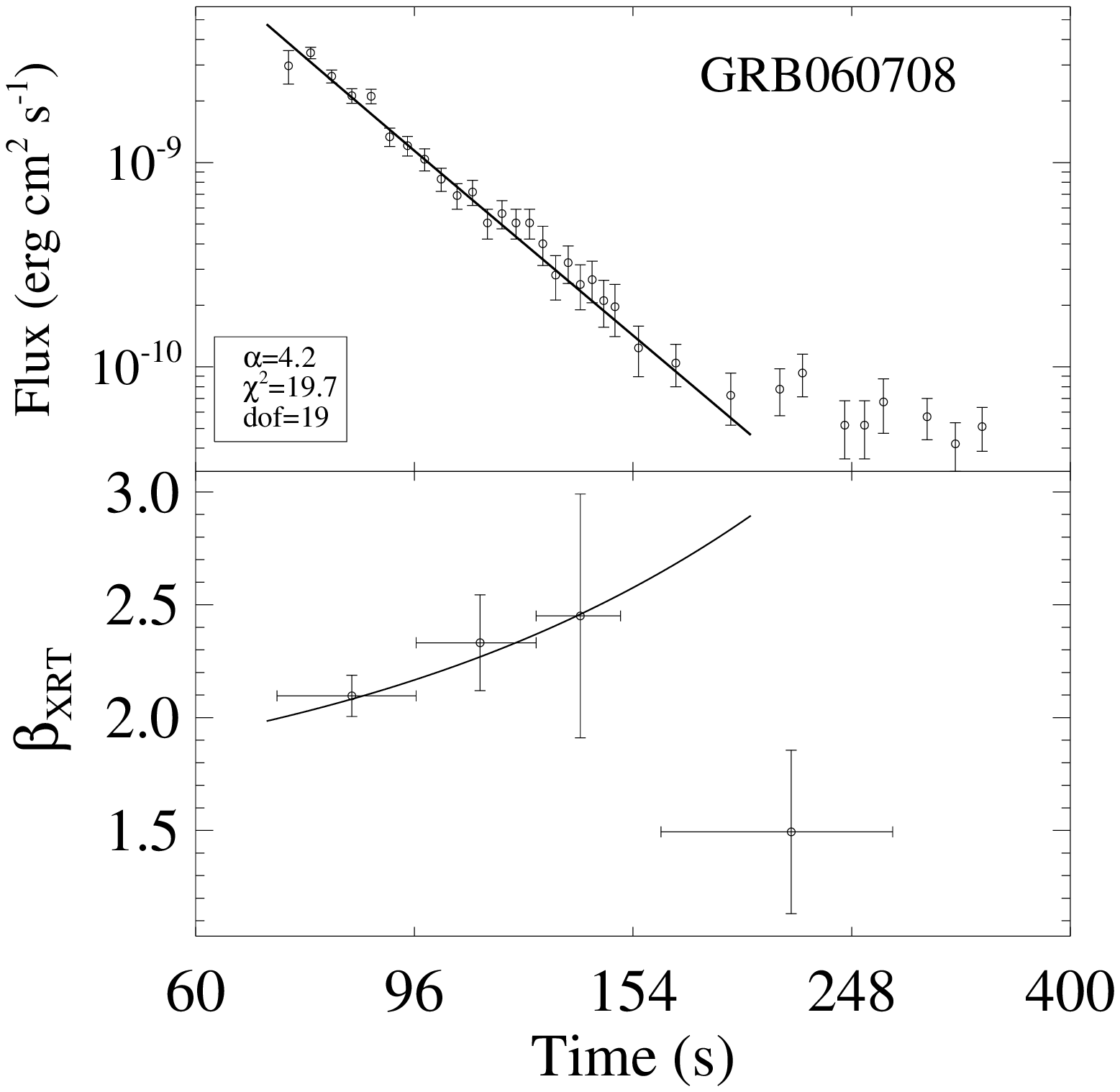}
\caption{Same as Figure 1 but for those tails with significant spectral
evolution but without superposing strong flares (Group B). The solid lines
show the results of our proposed modeling. }
\end{figure}

\clearpage

\begin{figure}
\includegraphics[angle=0,scale=.32]{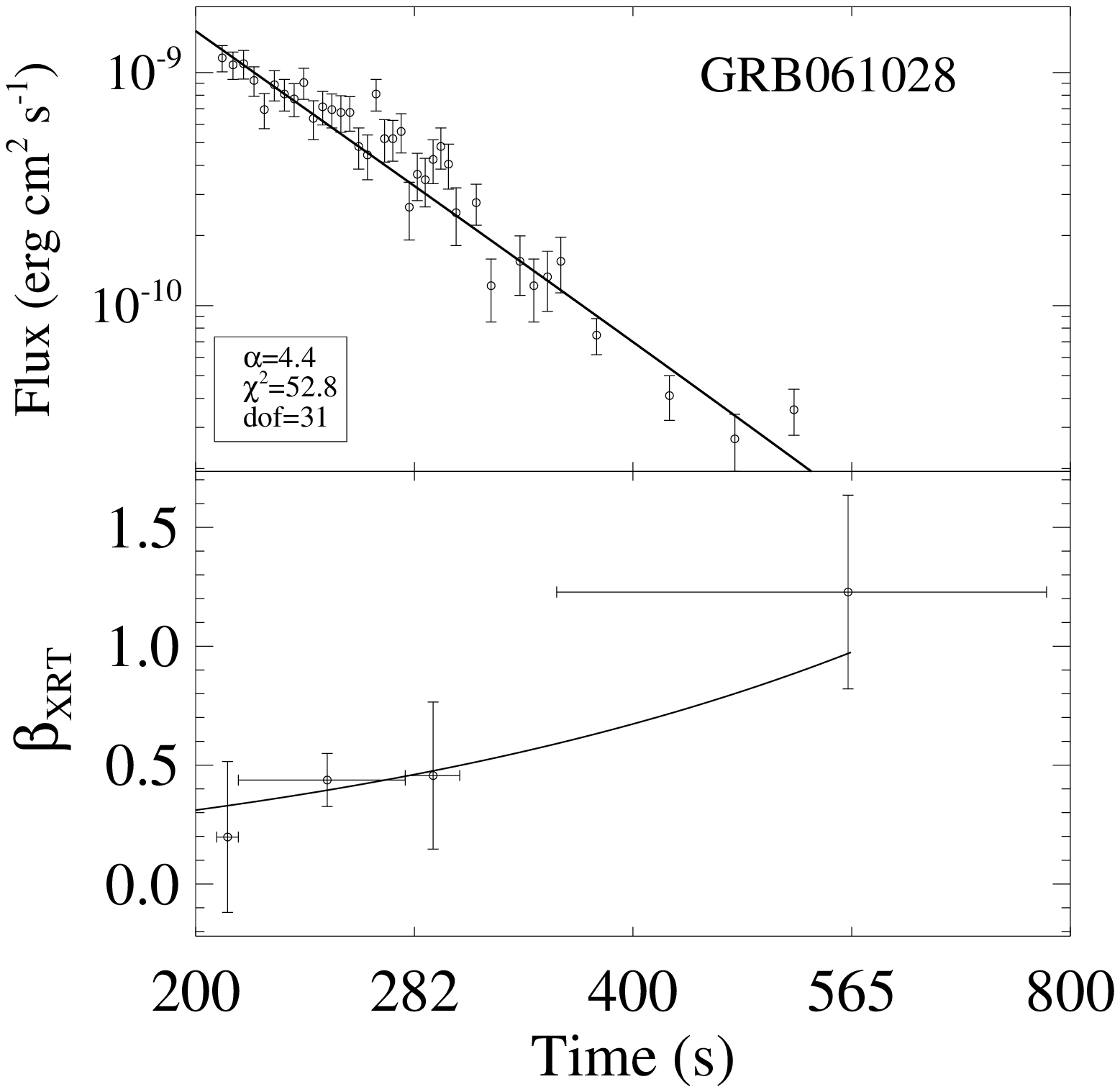}
\includegraphics[angle=0,scale=.32]{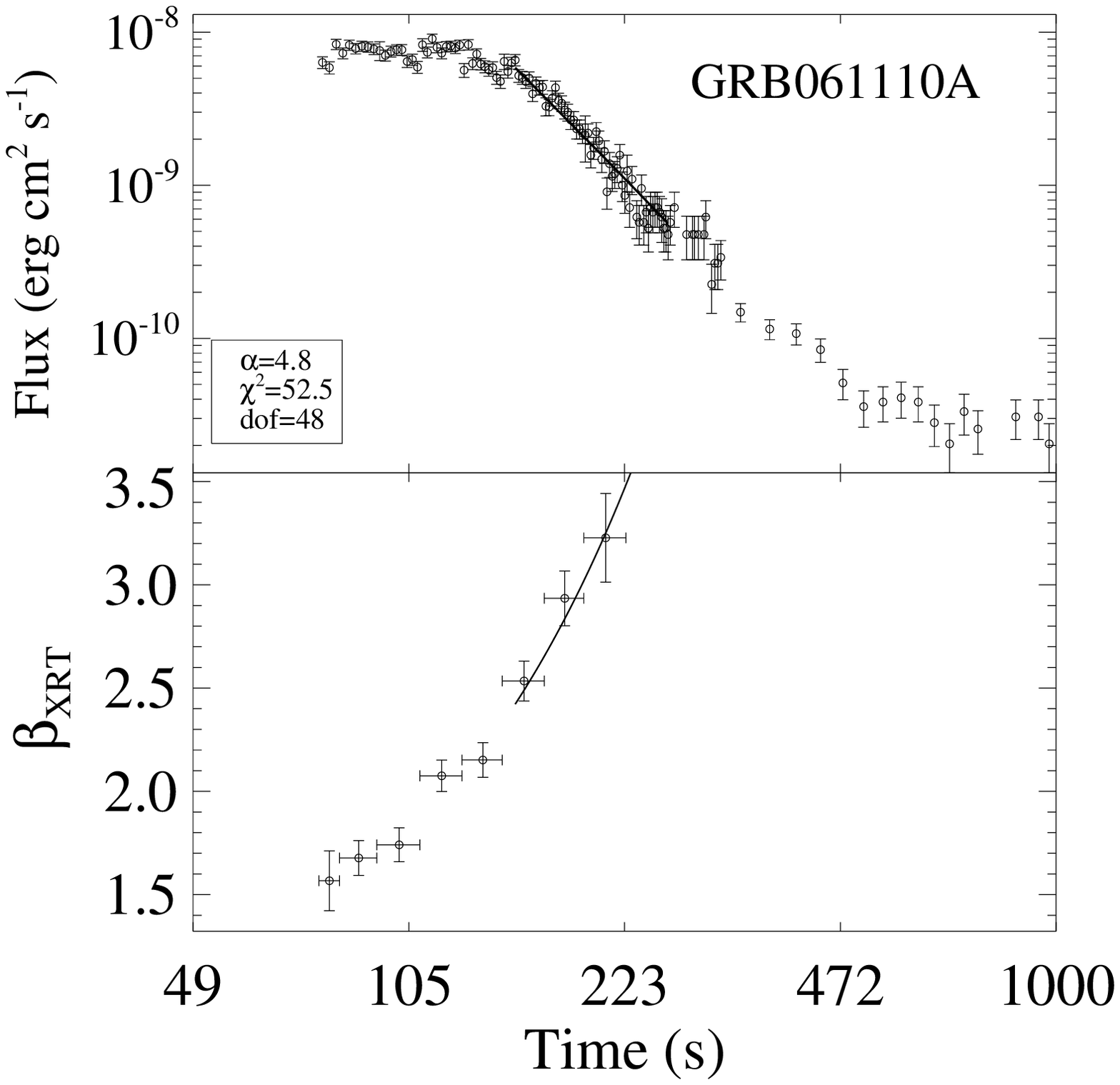}
\includegraphics[angle=0,scale=.32]{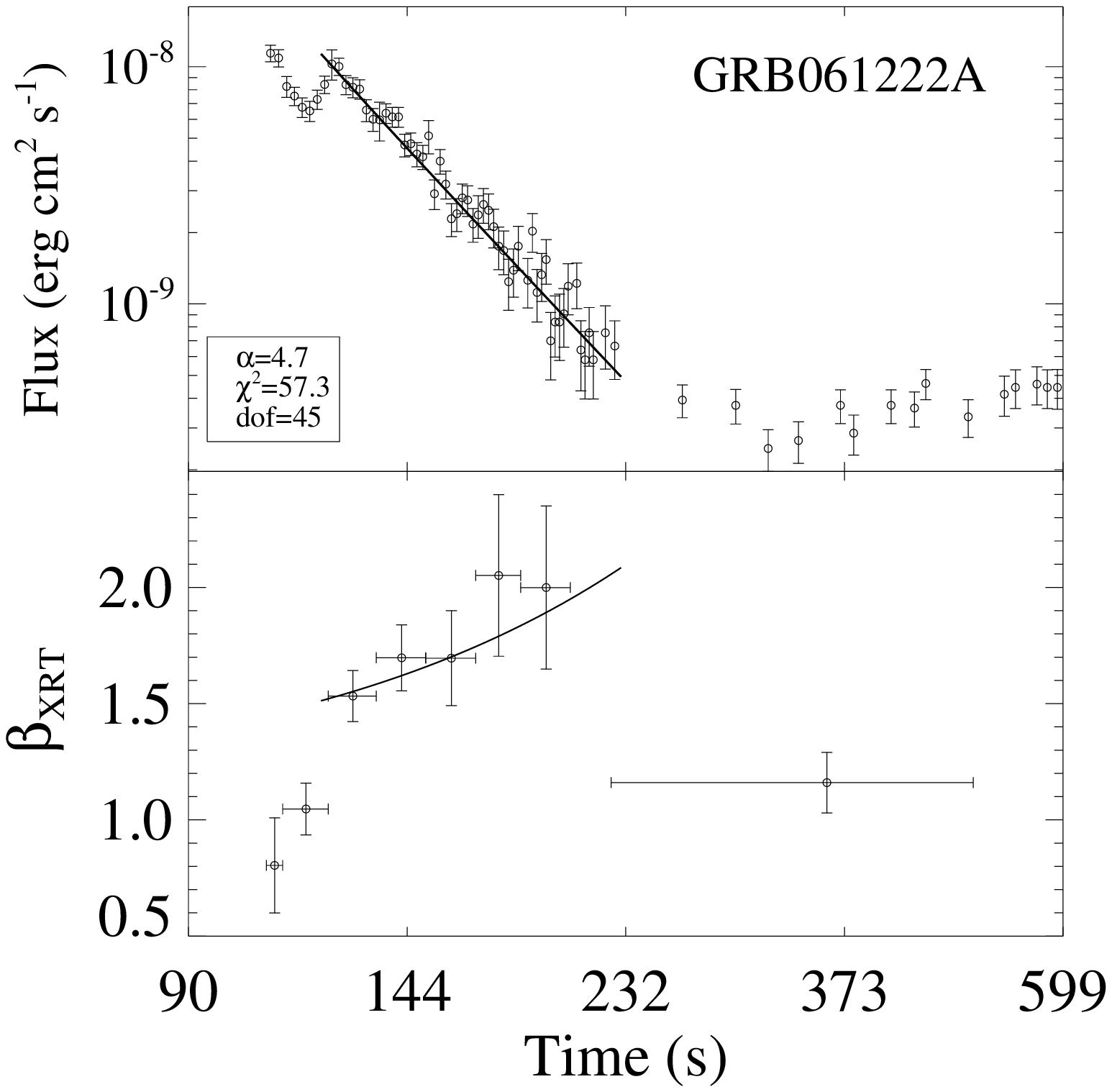}
  \hfill
\includegraphics[angle=0,scale=.32]{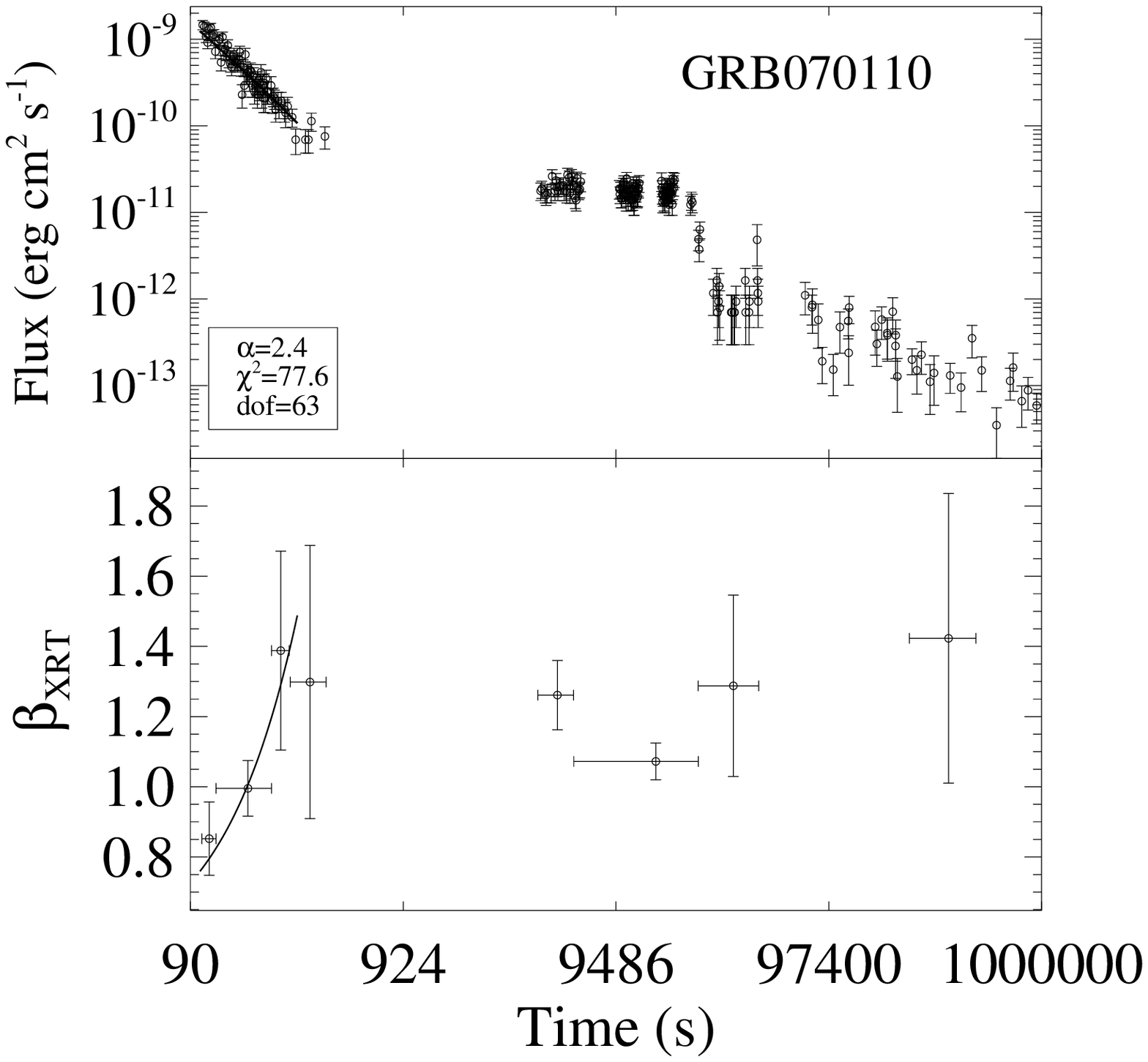}
\hfill
\hfill
\hfill
\hfill
\hfill
\hfill
\hfill
\hfill
 \caption{Continued.}
\end{figure}

\clearpage

\renewcommand\thefigure{3}
\begin{figure}
\includegraphics[angle=0,scale=.32]{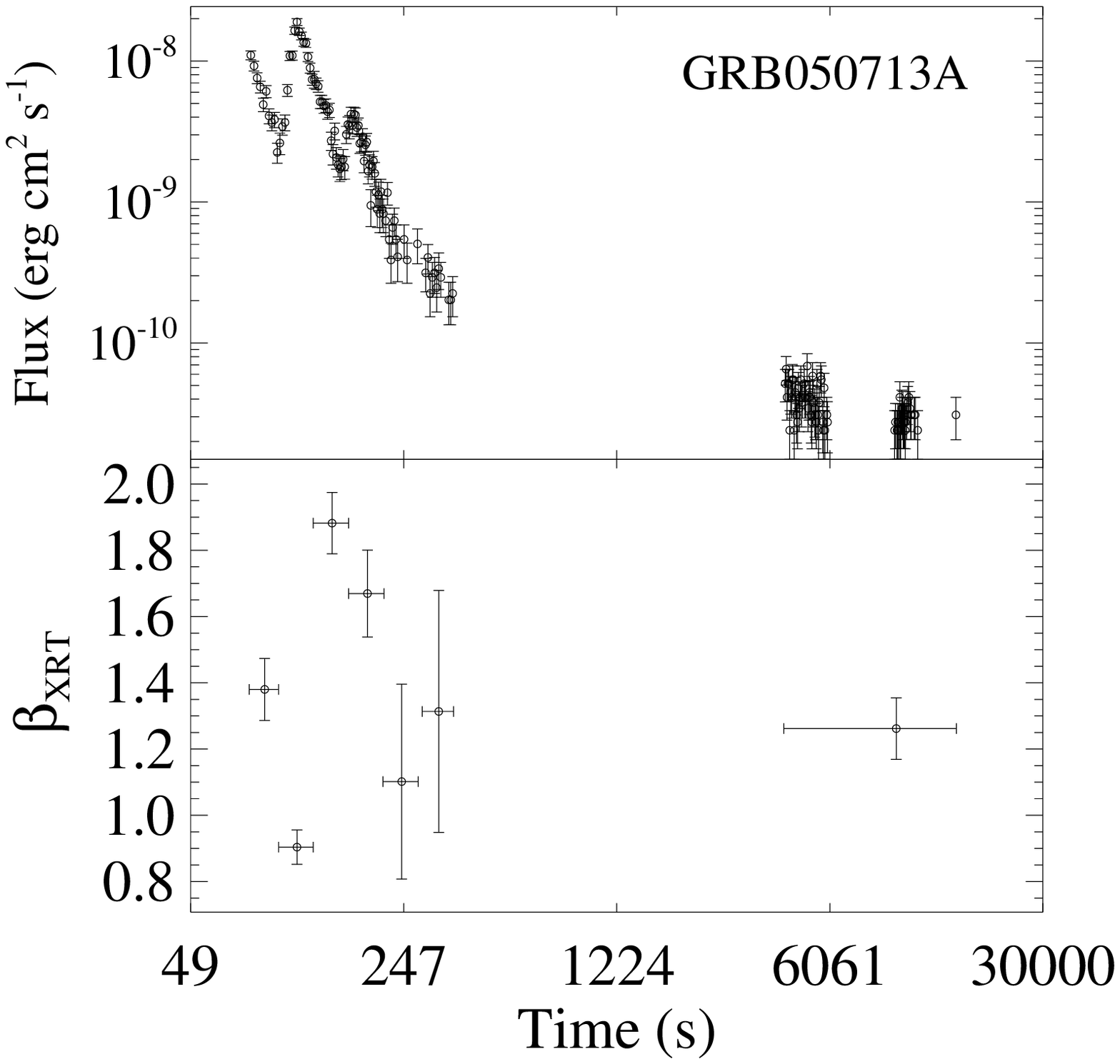}
\includegraphics[angle=0,scale=.32]{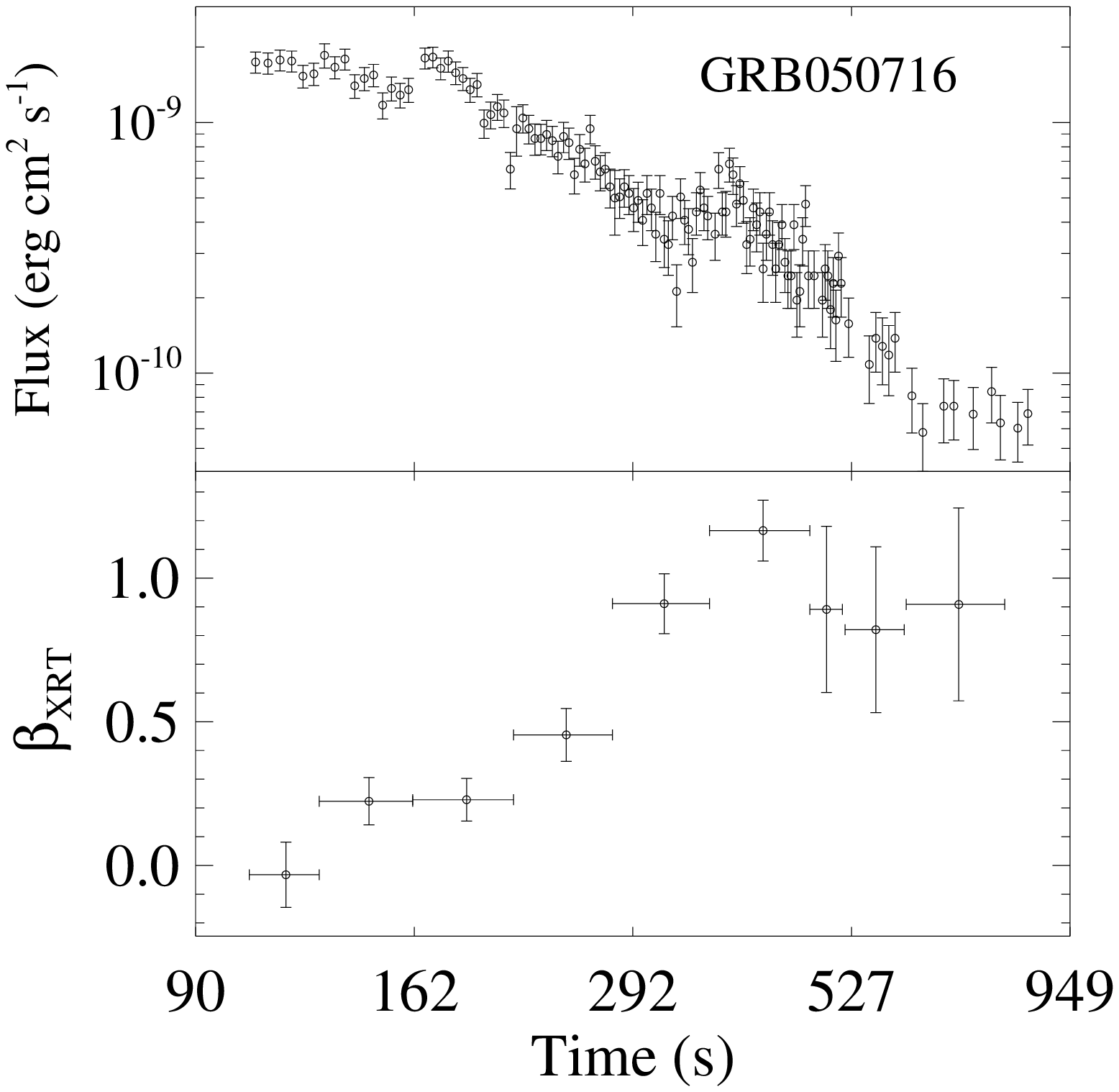}
\includegraphics[angle=0,scale=.32]{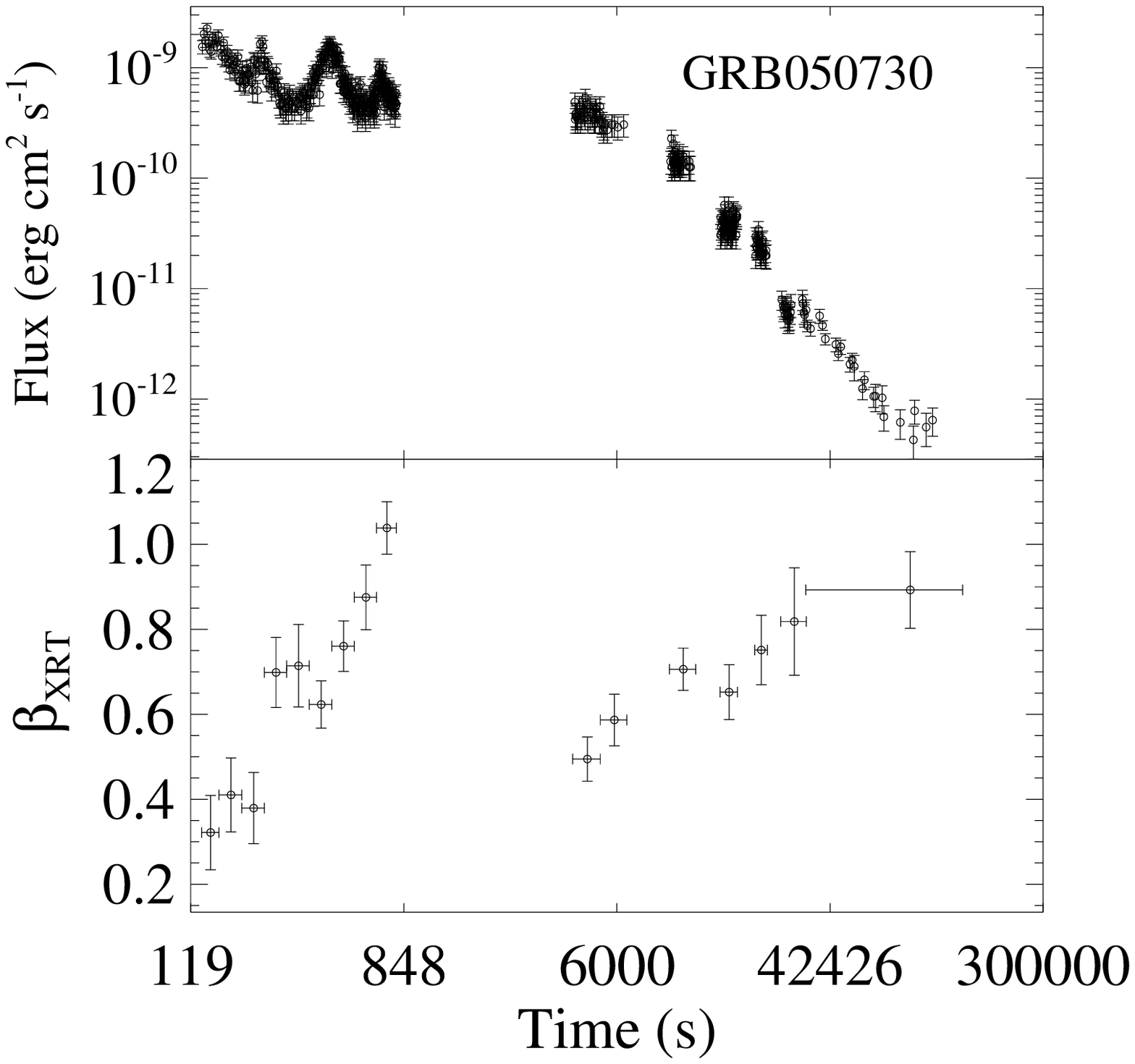}
\hfill
\includegraphics[angle=0,scale=.32]{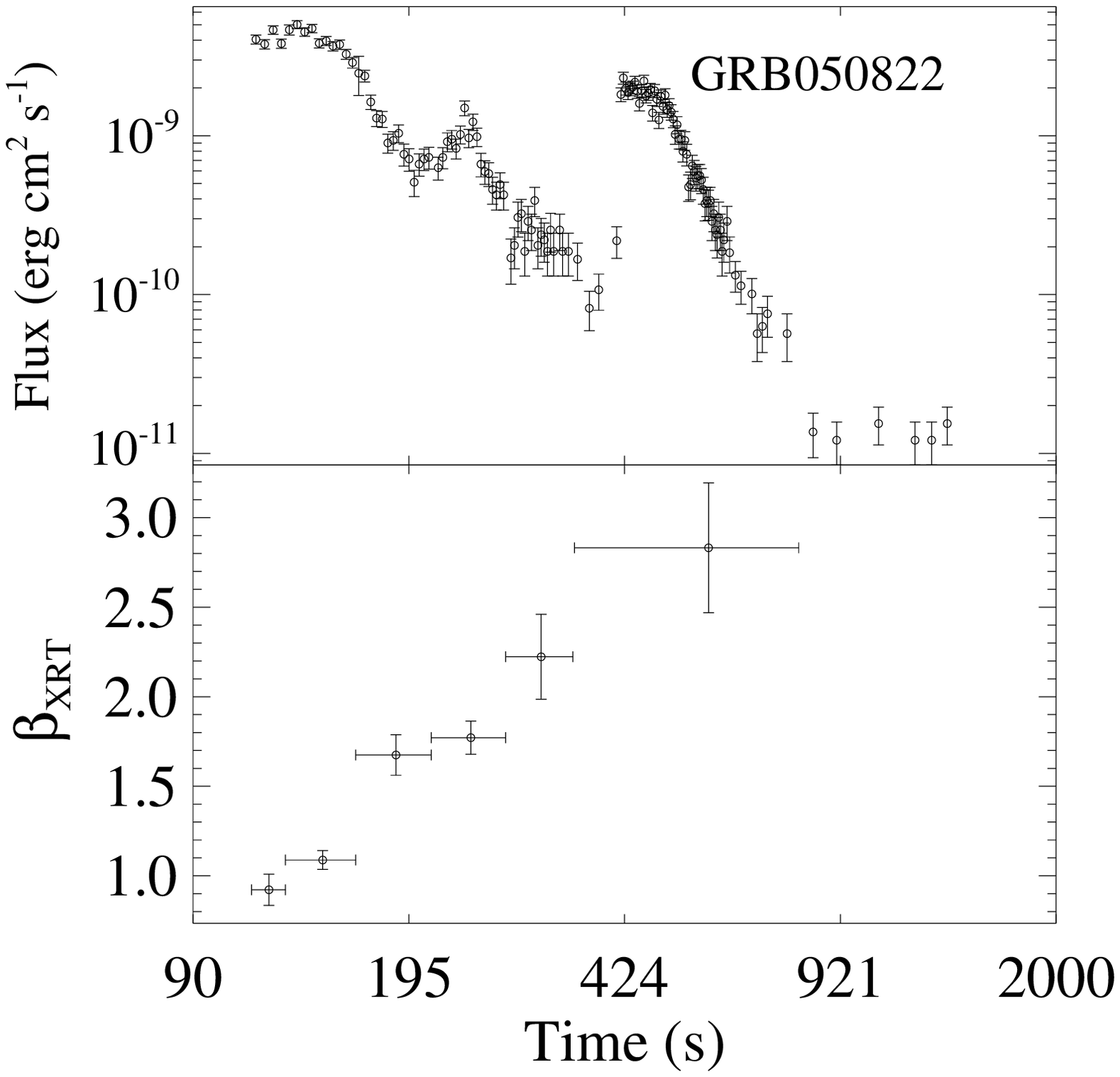}
\includegraphics[angle=0,scale=.32]{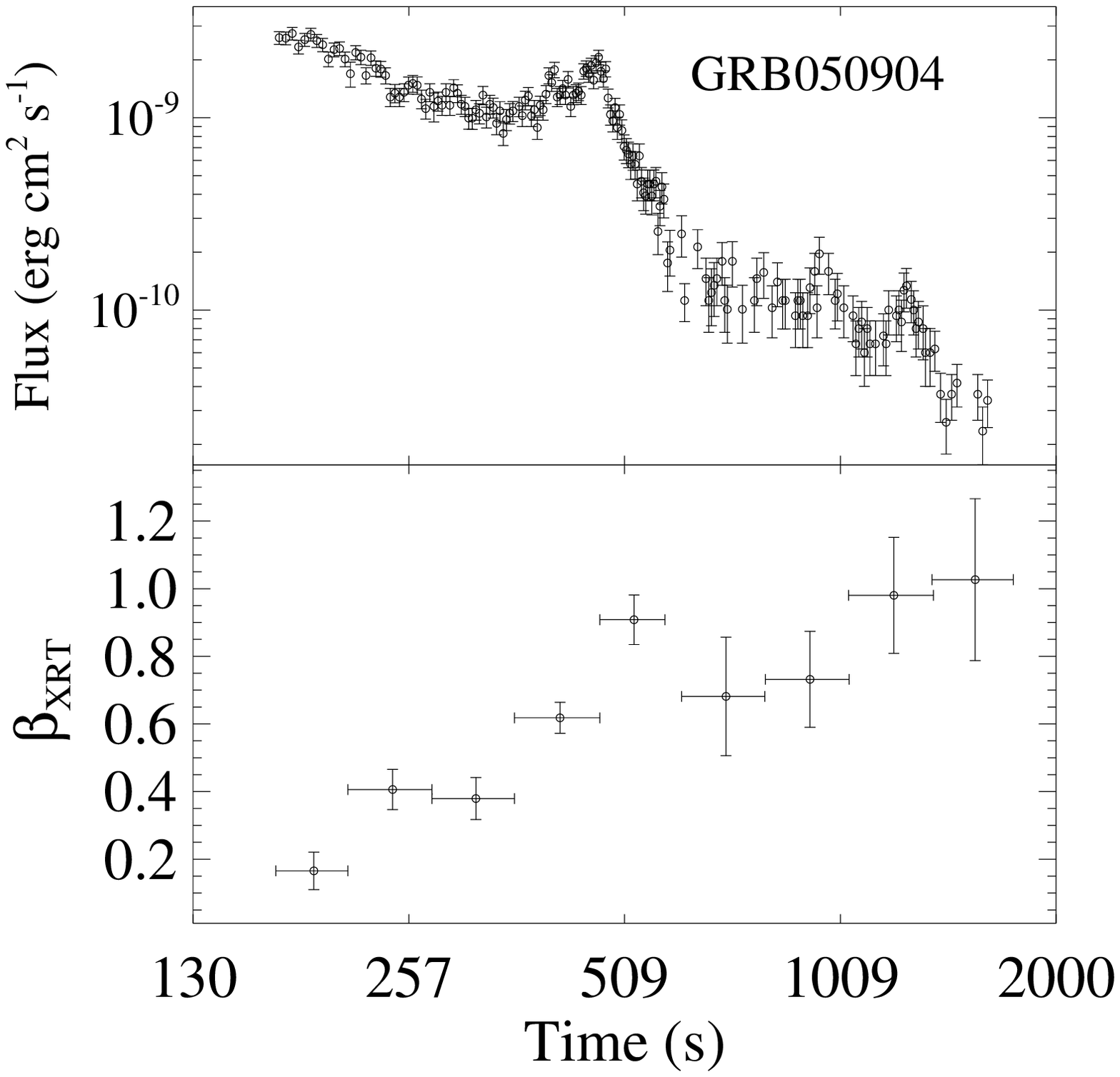}
\includegraphics[angle=0,scale=.32]{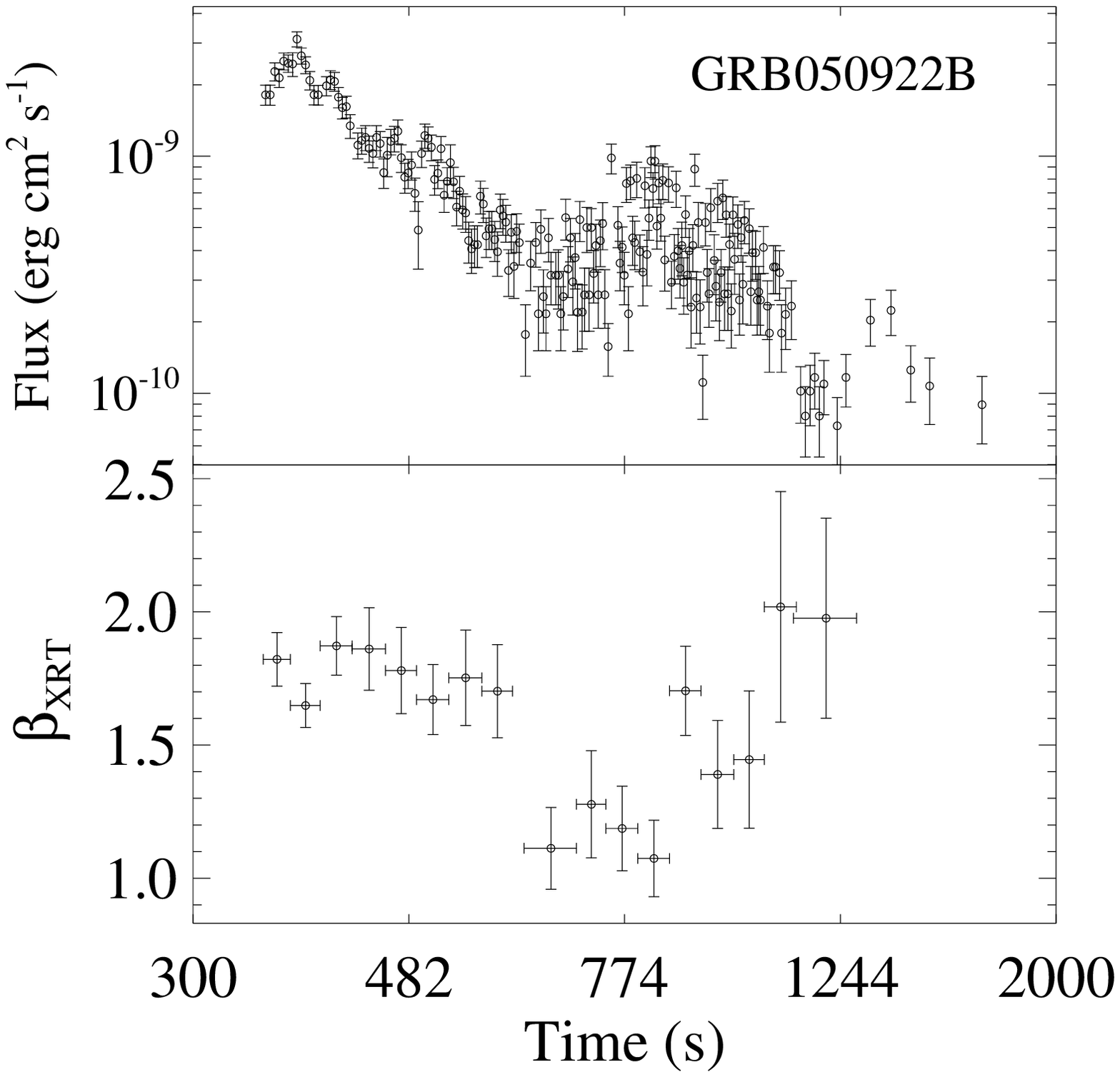}
\hfill
\includegraphics[angle=0,scale=.32]{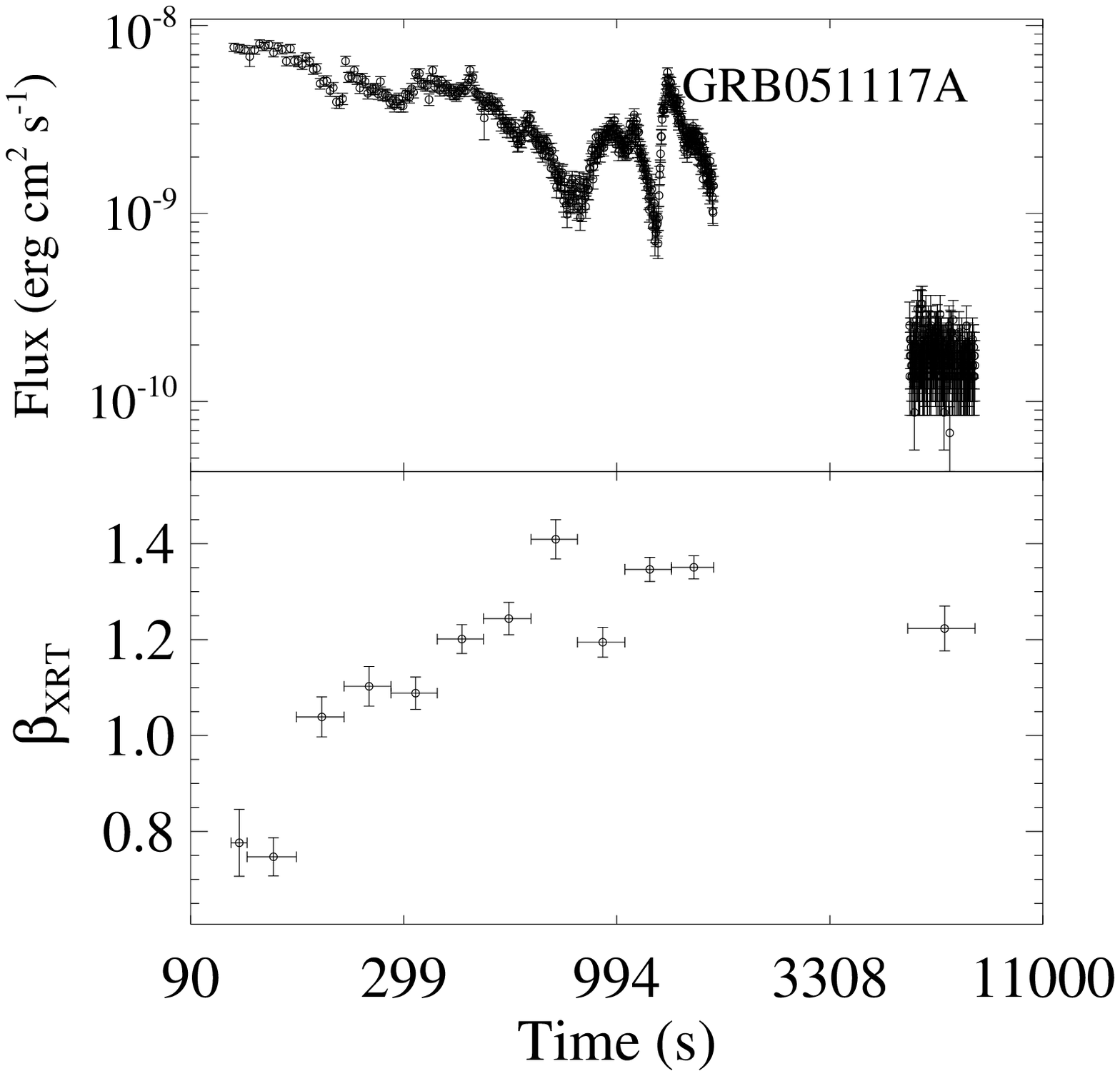}
\includegraphics[angle=0,scale=.32]{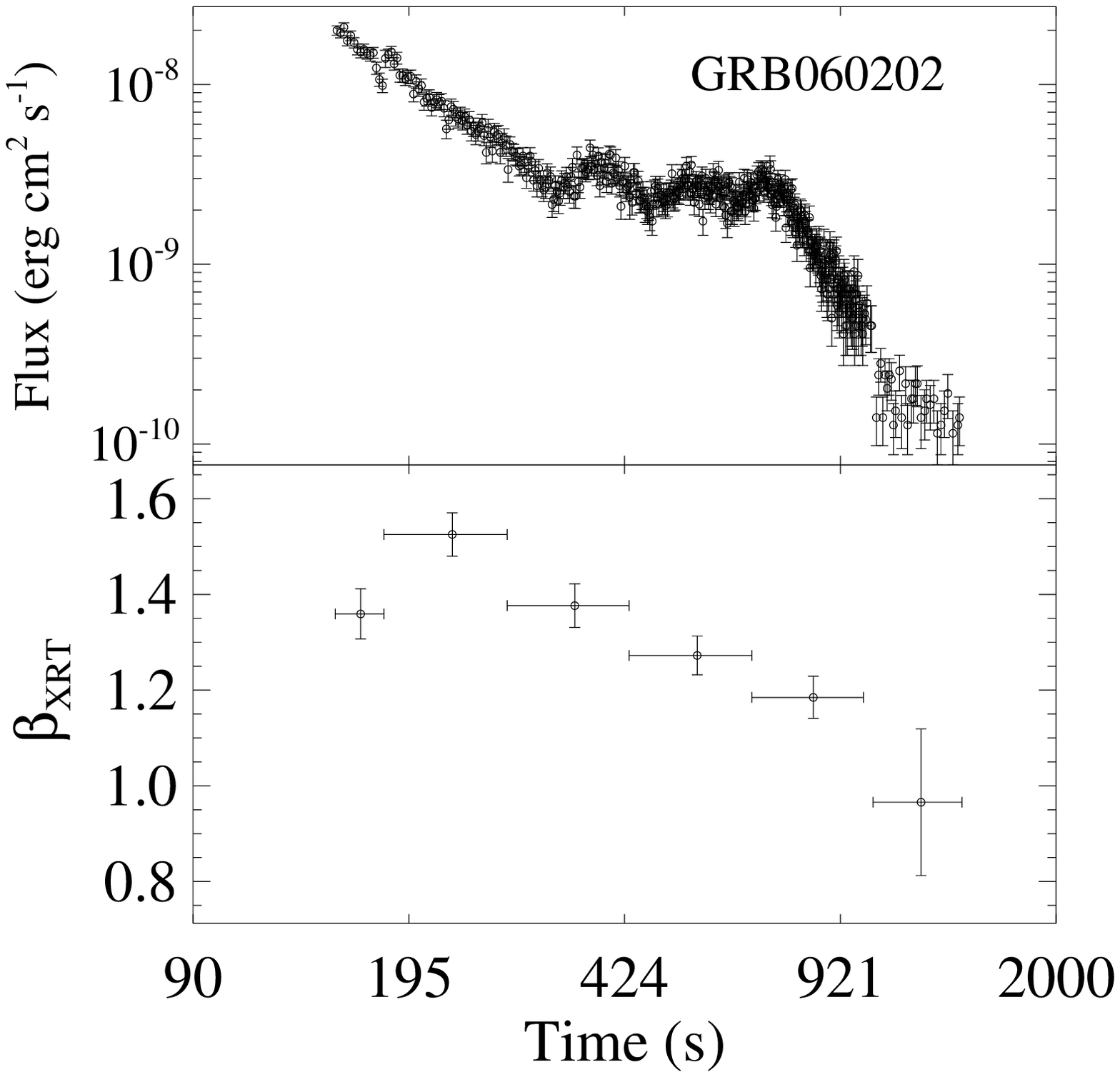}
\includegraphics[angle=0,scale=.32]{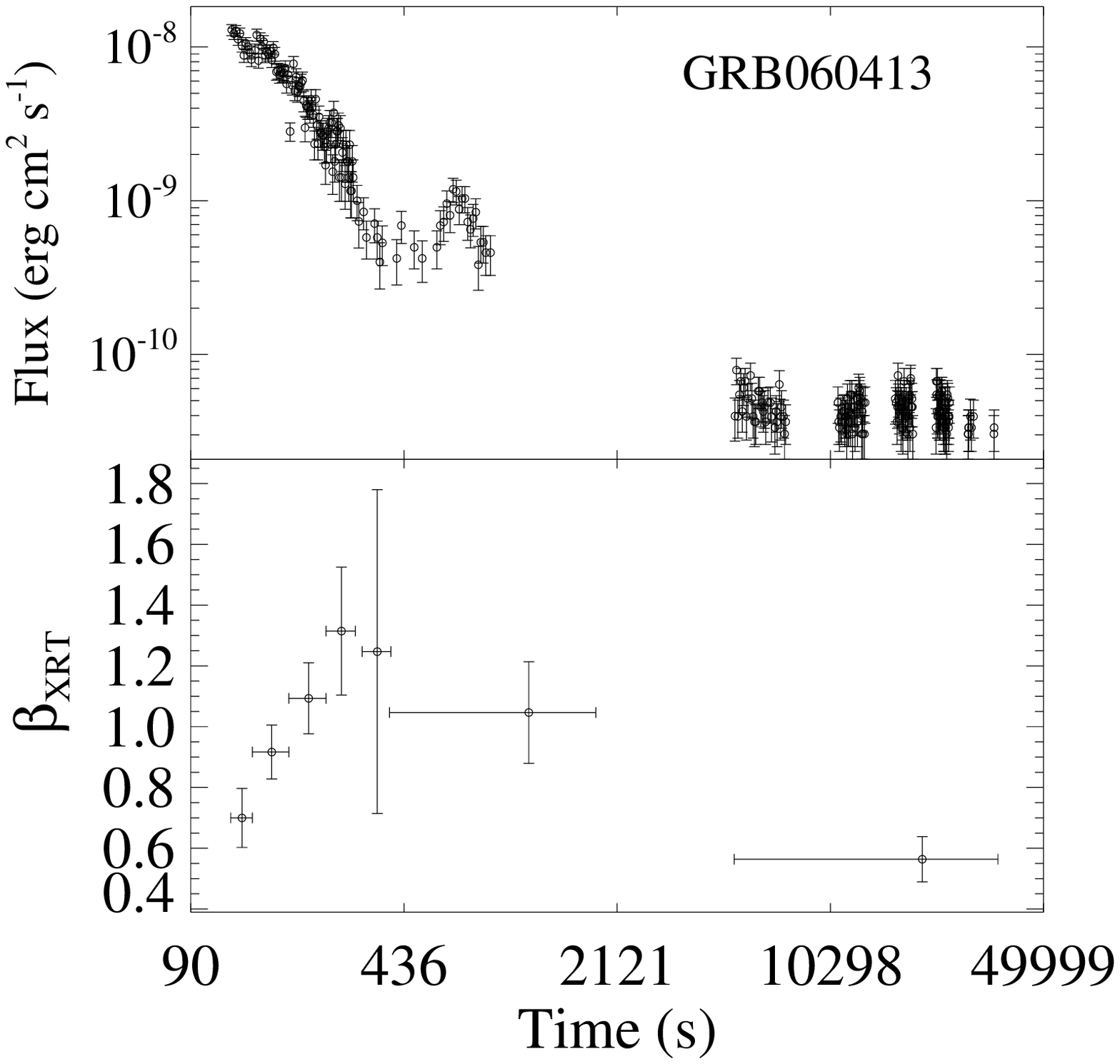}
\hfill
\includegraphics[angle=0,scale=.32]{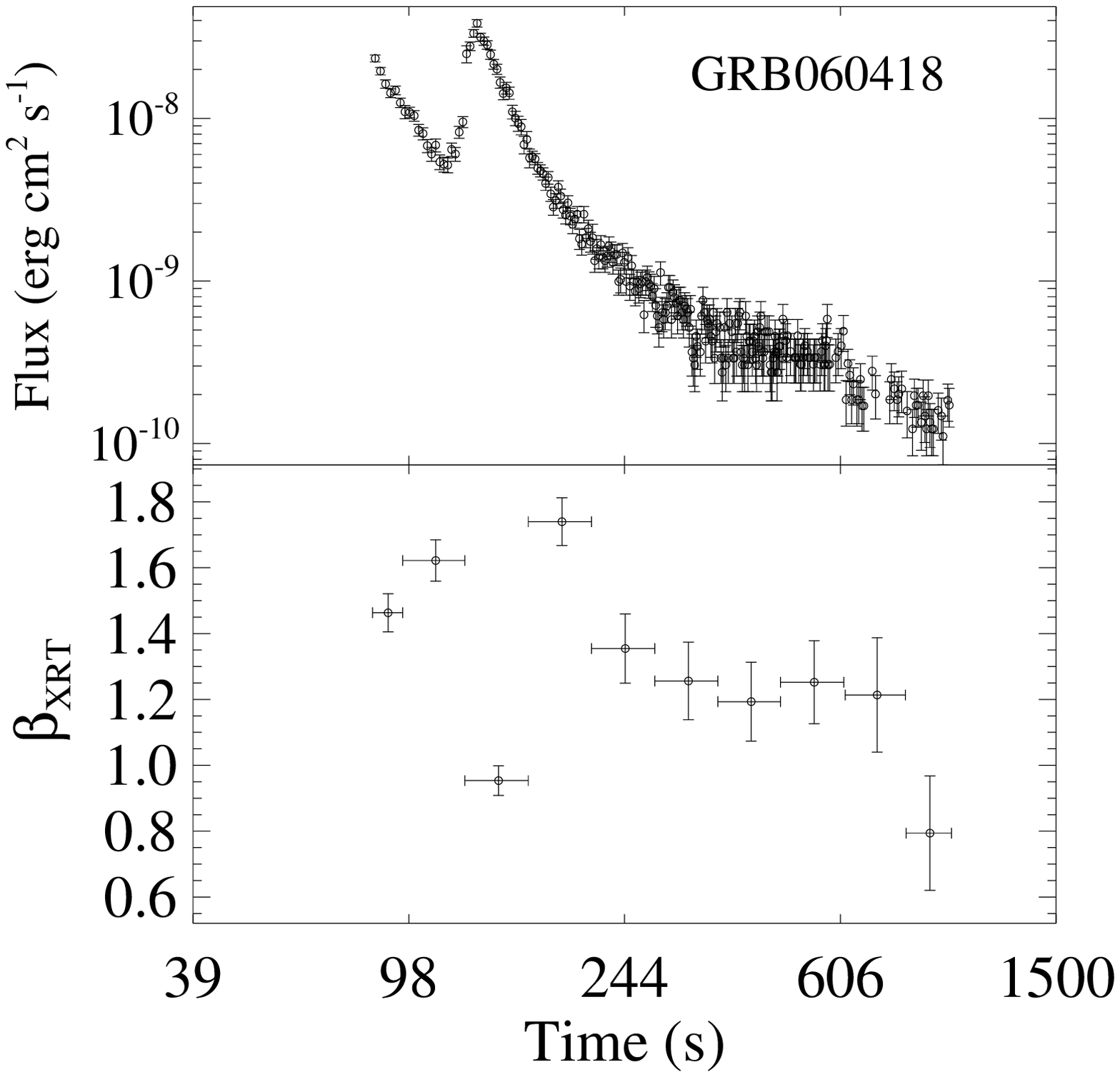}
\hfill
\includegraphics[angle=0,scale=.32]{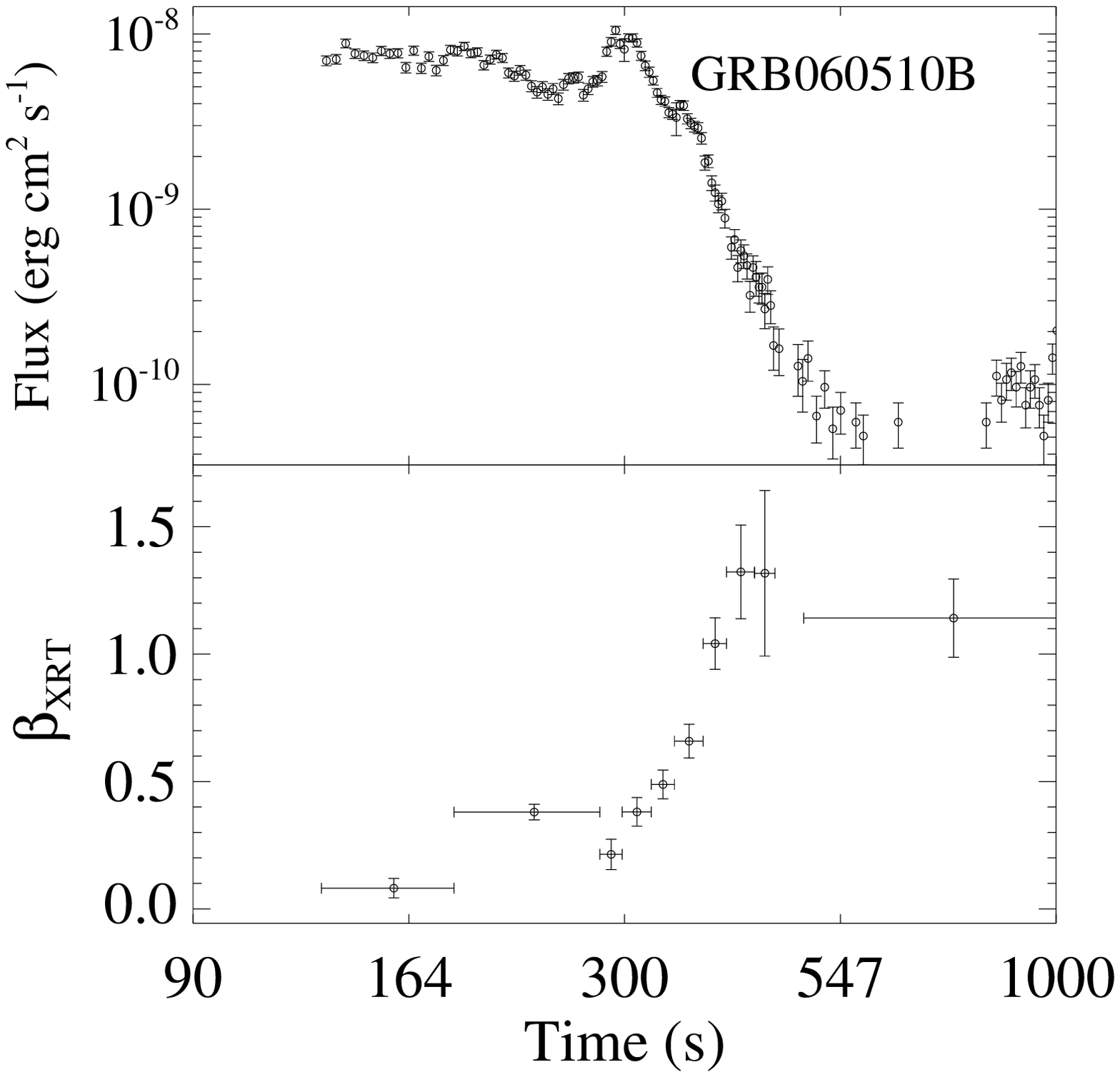}
\hfill
\includegraphics[angle=0,scale=.32]{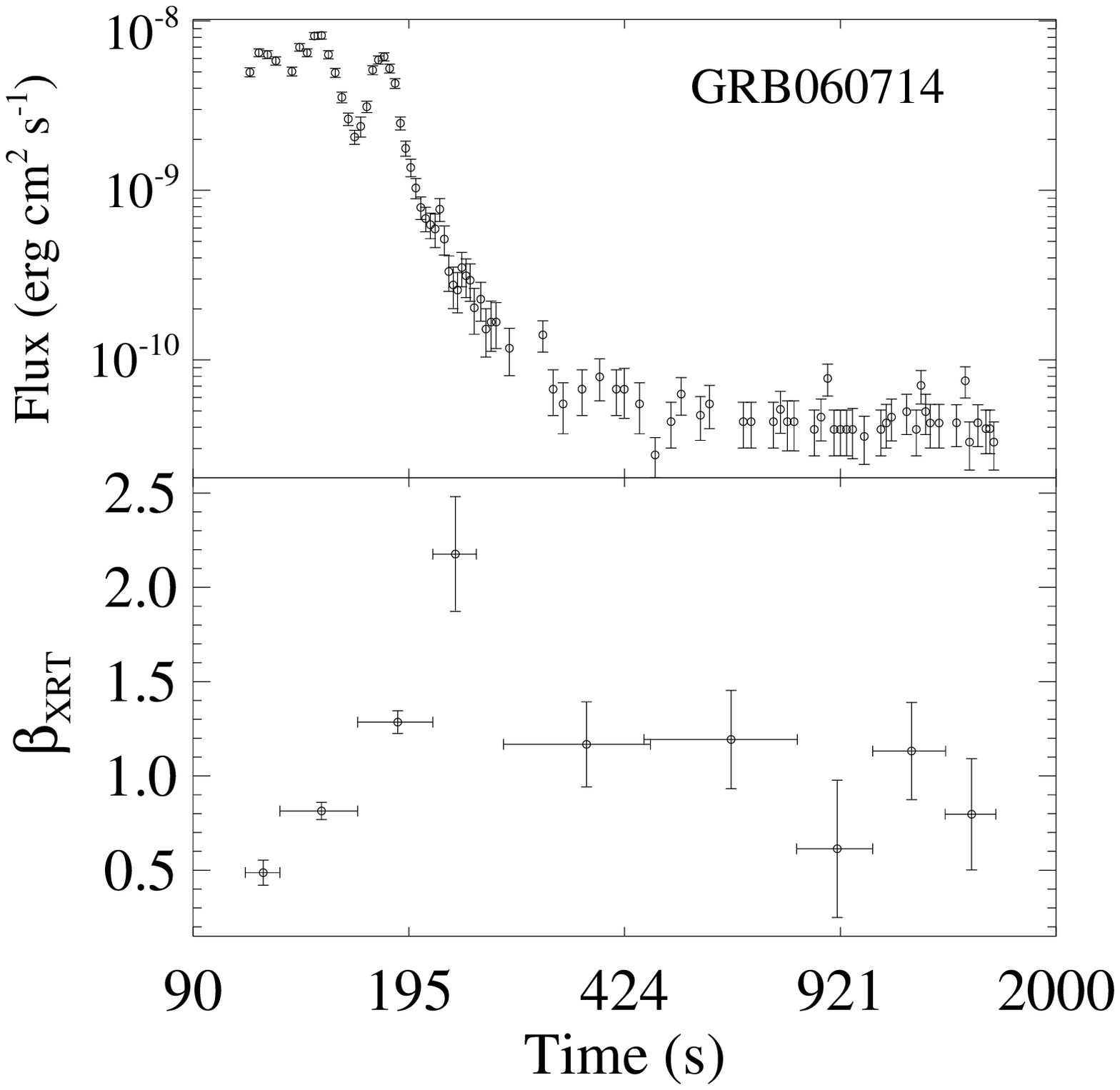}
\caption{Same as Figure 1 but for those tails with significant flare contamination
(Group C).}
\end{figure}

\clearpage

\begin{figure}
\includegraphics[angle=0,scale=.32]{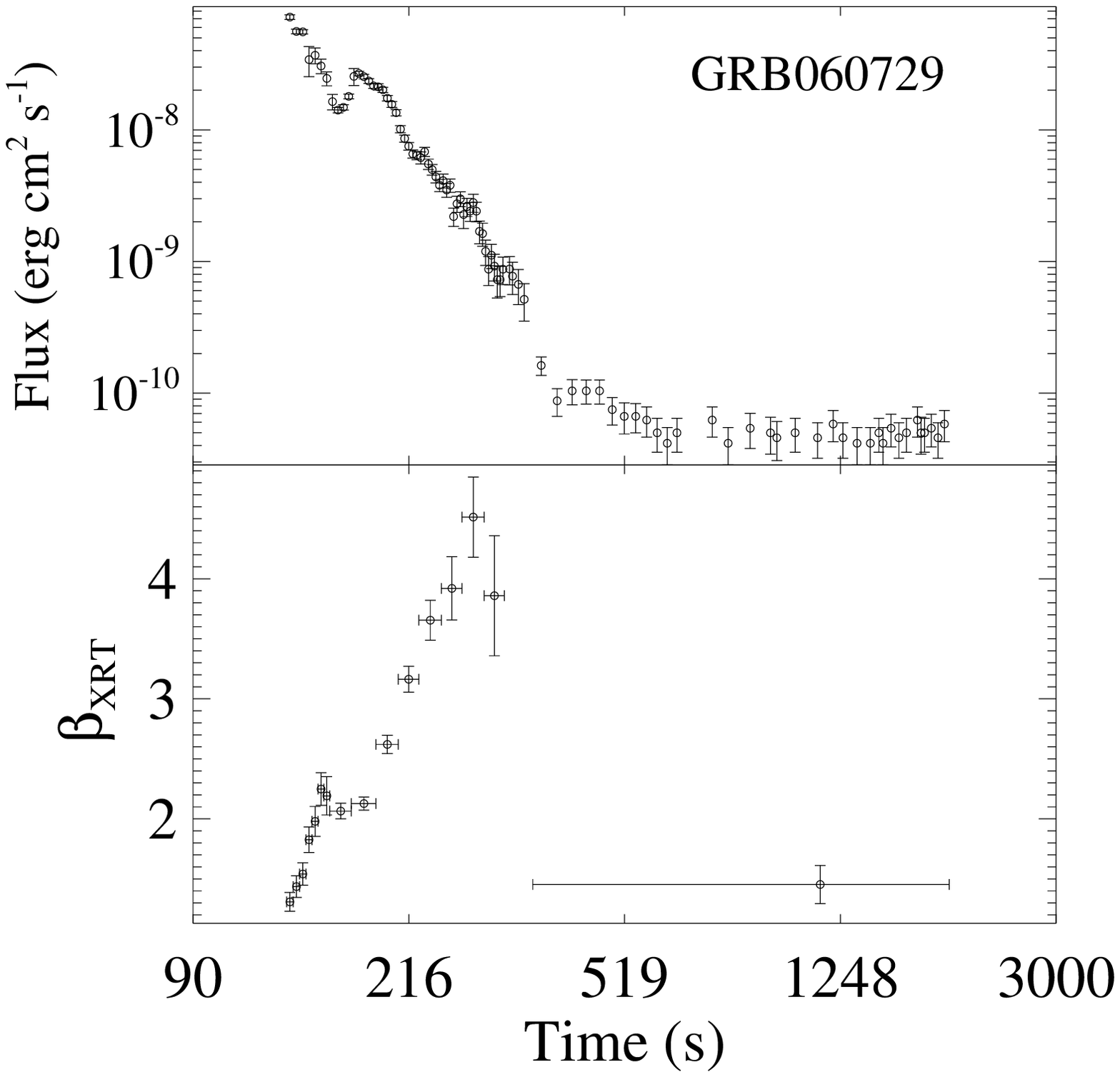}
\includegraphics[angle=0,scale=.32]{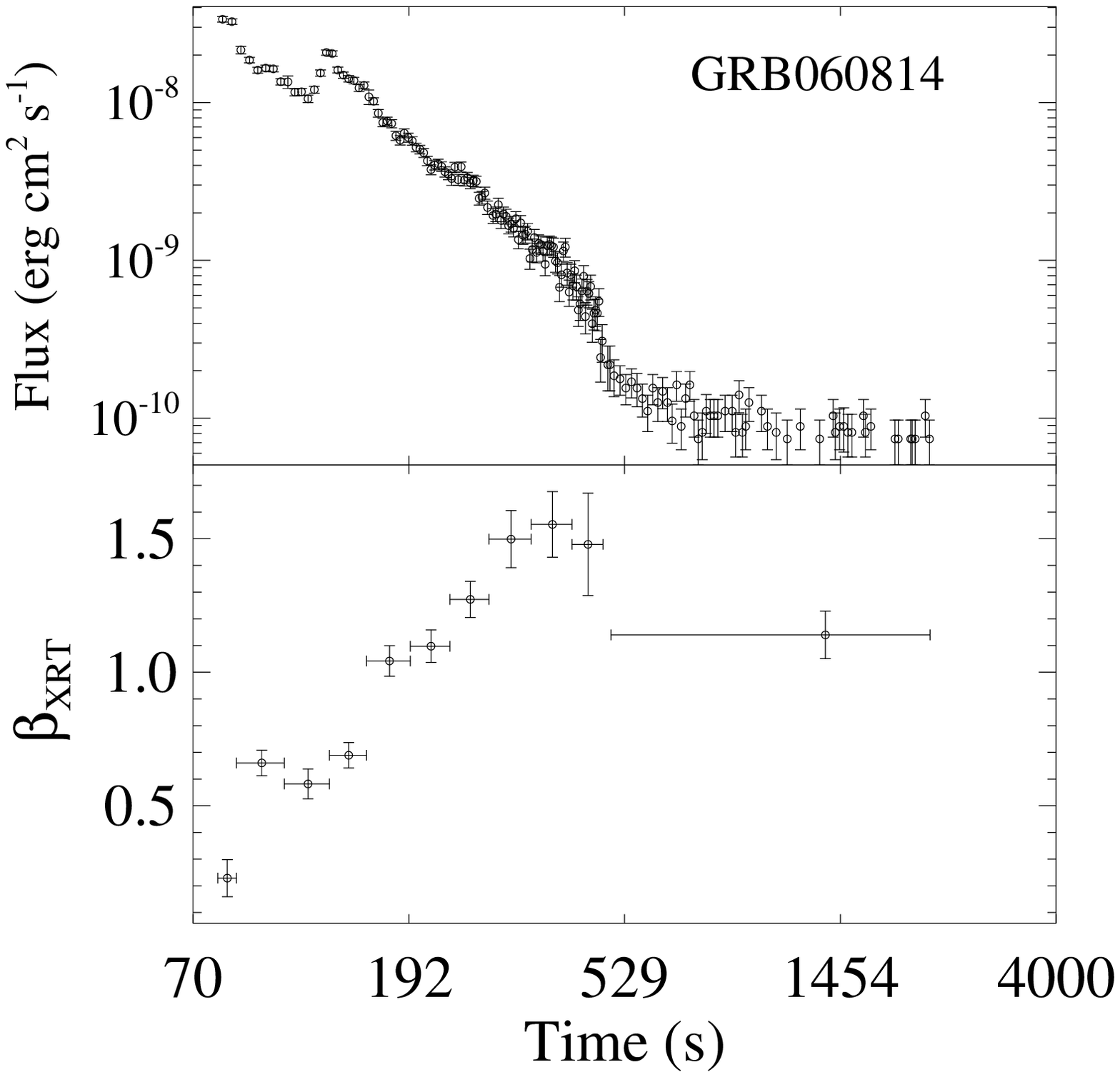}
\includegraphics[angle=0,scale=.32]{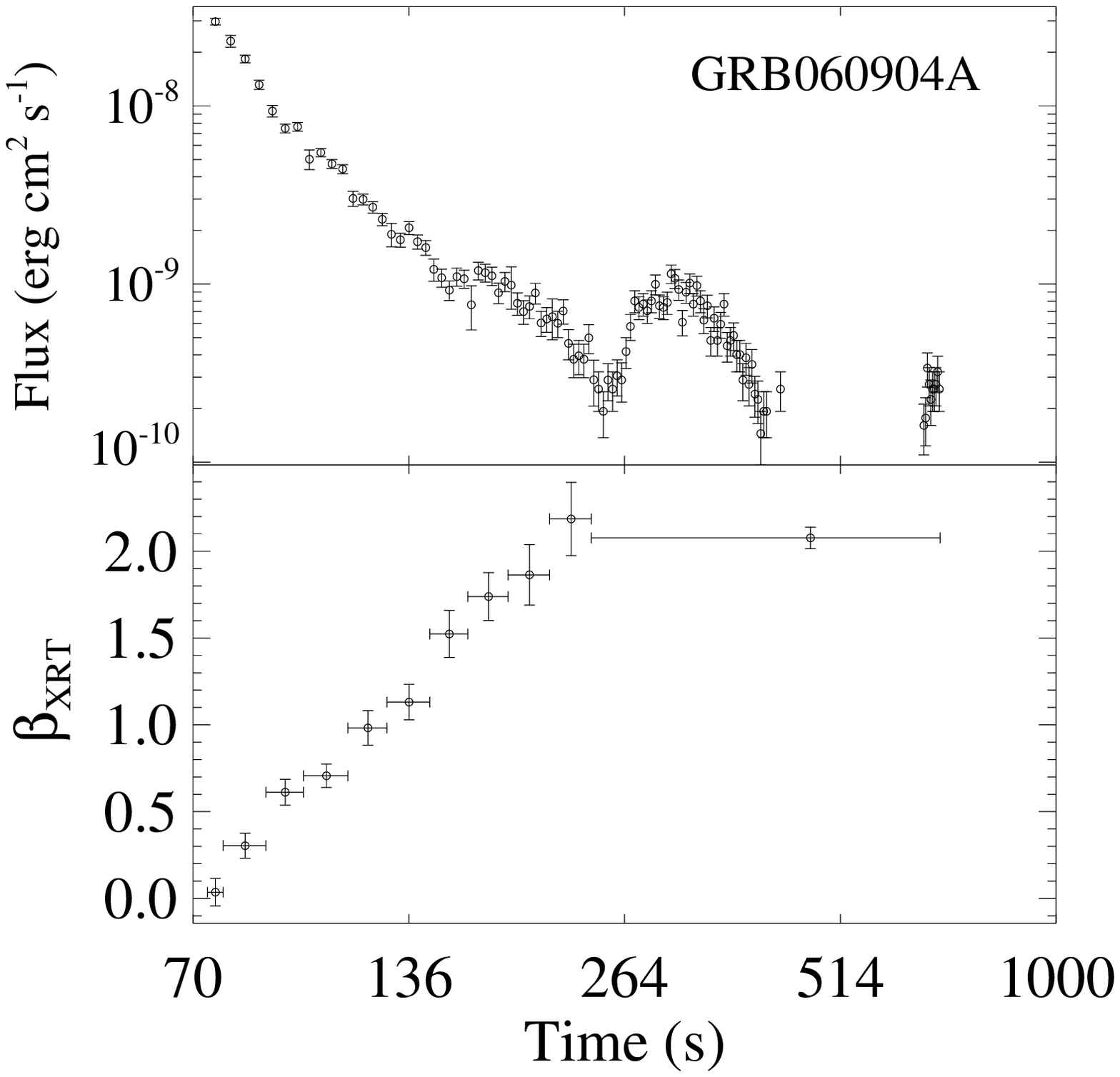}
\hfill
\includegraphics[angle=0,scale=.32]{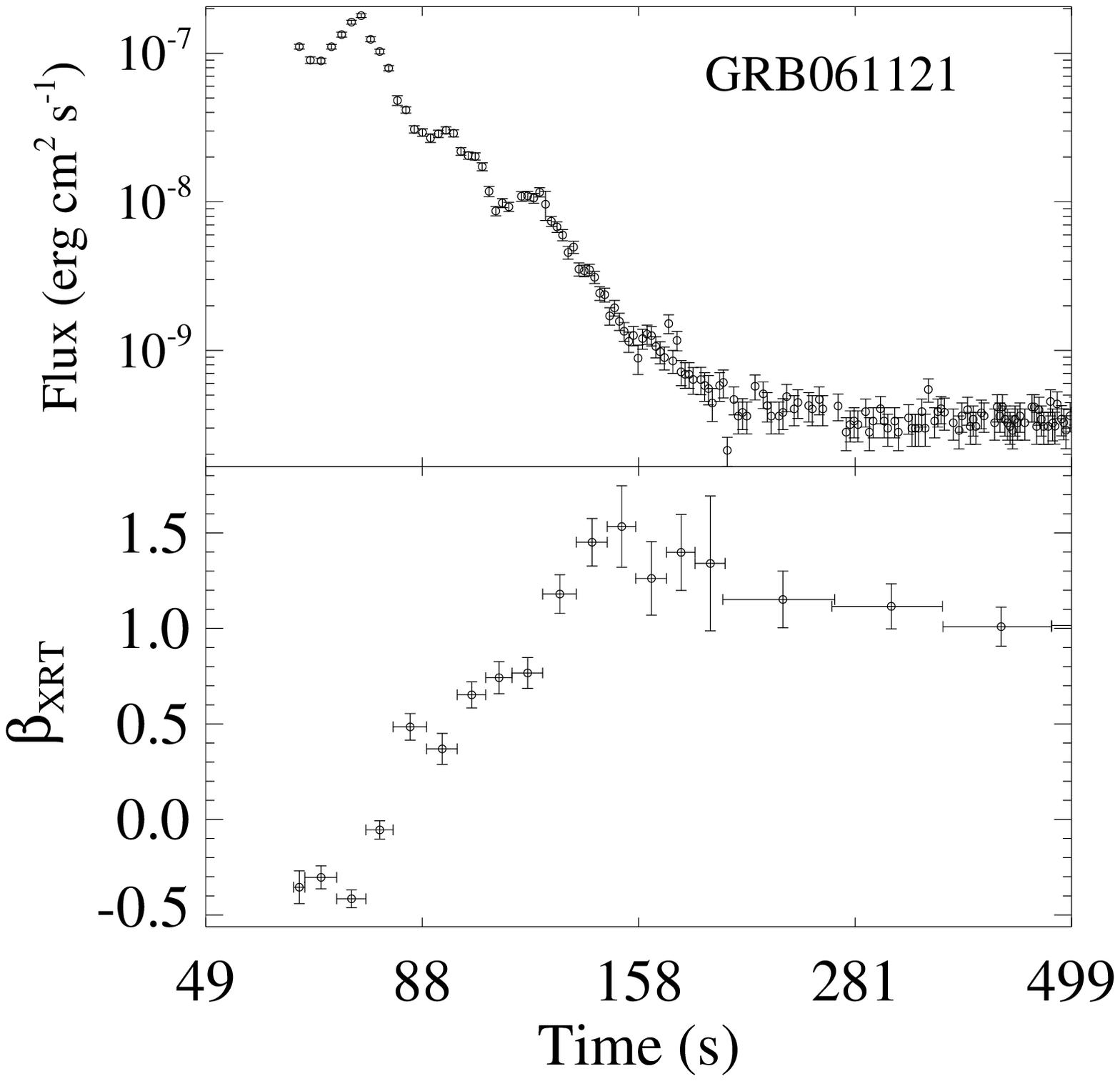}
\hfill
\includegraphics[angle=0,scale=.32]{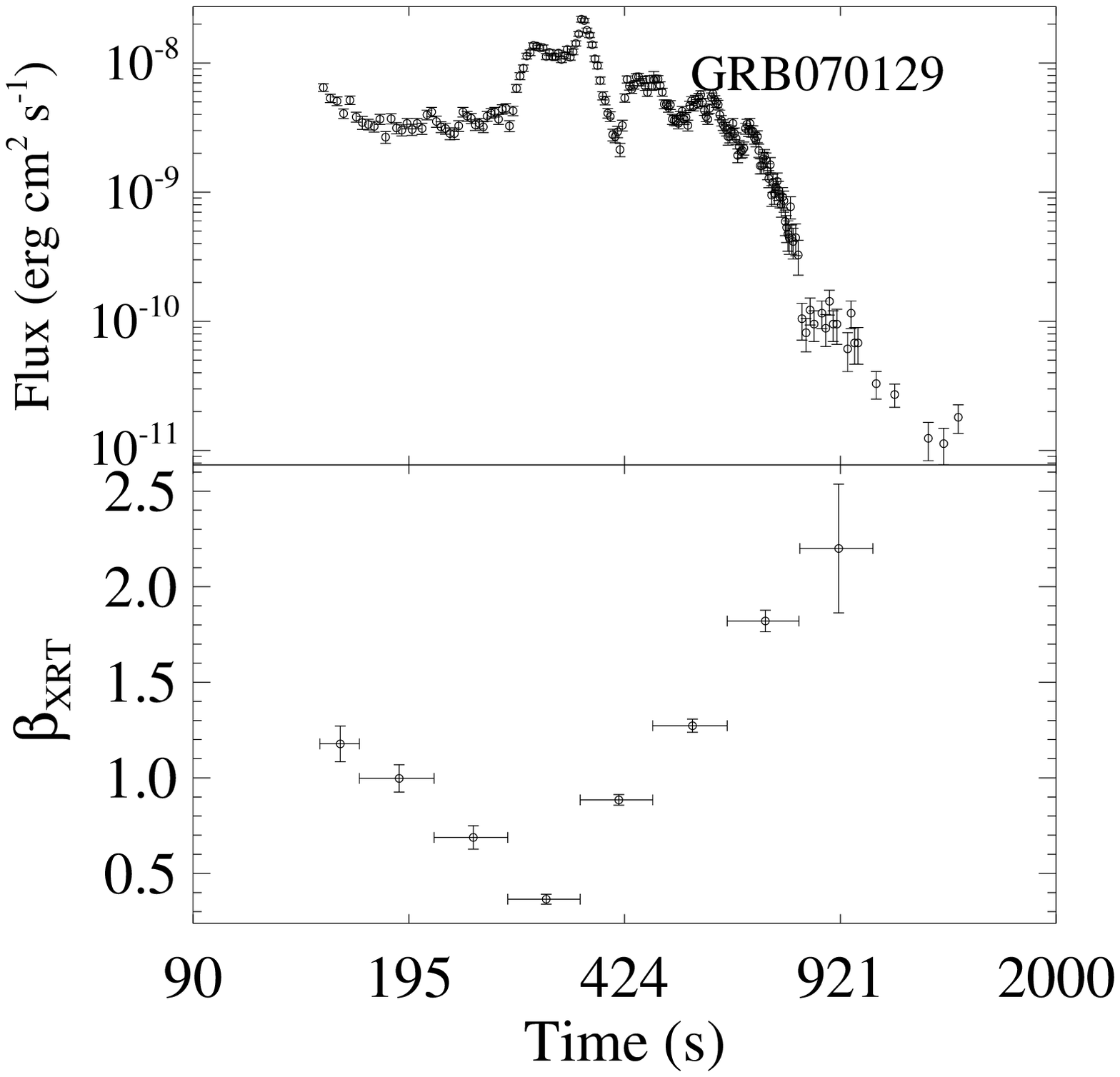}
\hfill
\hfill
\hfill
\hfill
\hfill
\hfill
\hfill
\hfill
\caption{Continued.}
\end{figure}

\clearpage

\renewcommand\thefigure{4}
\begin{figure}
\includegraphics[angle=0,width=3.2in,height=3.9in]{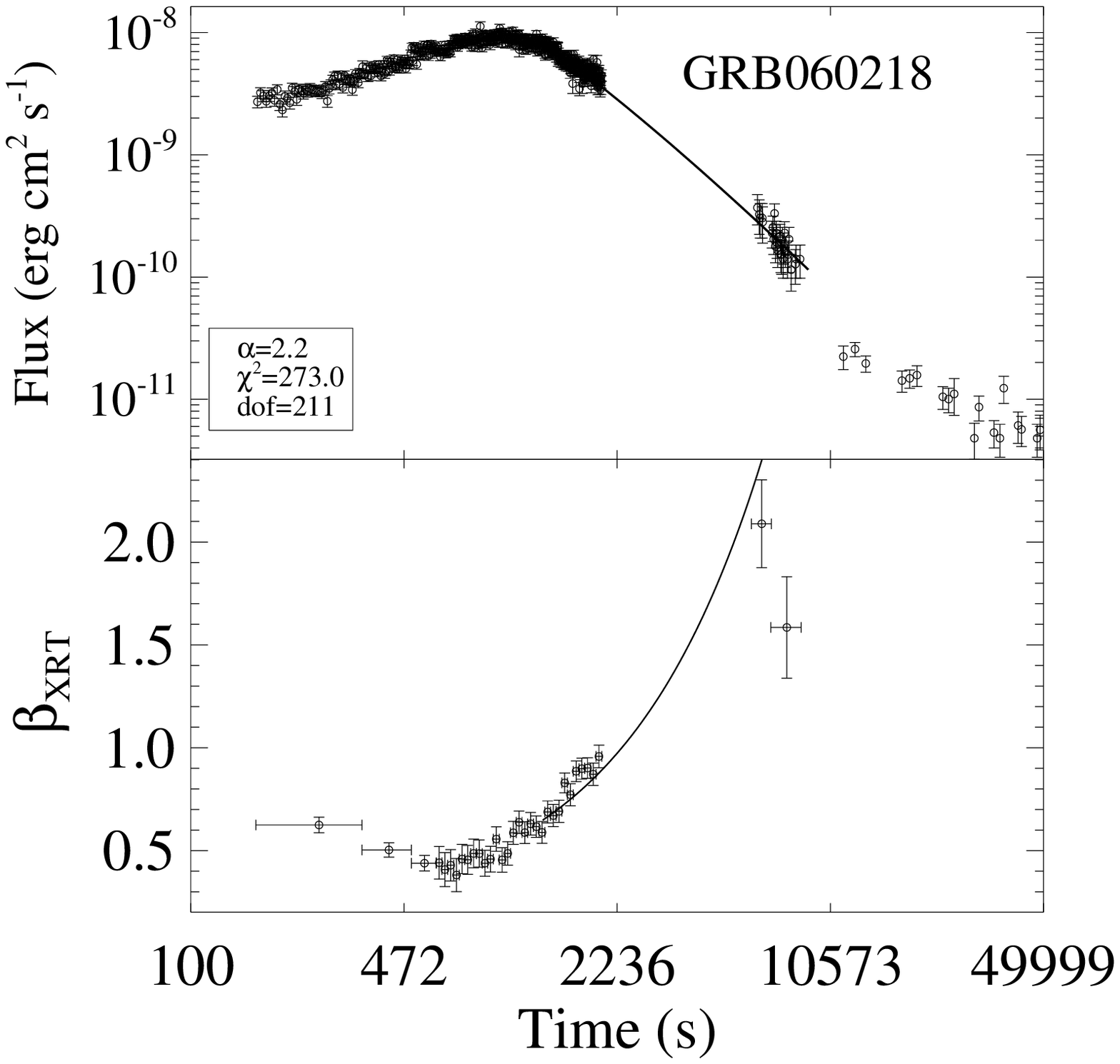}%
\includegraphics[angle=0,width=2.9in,height=4.3in]{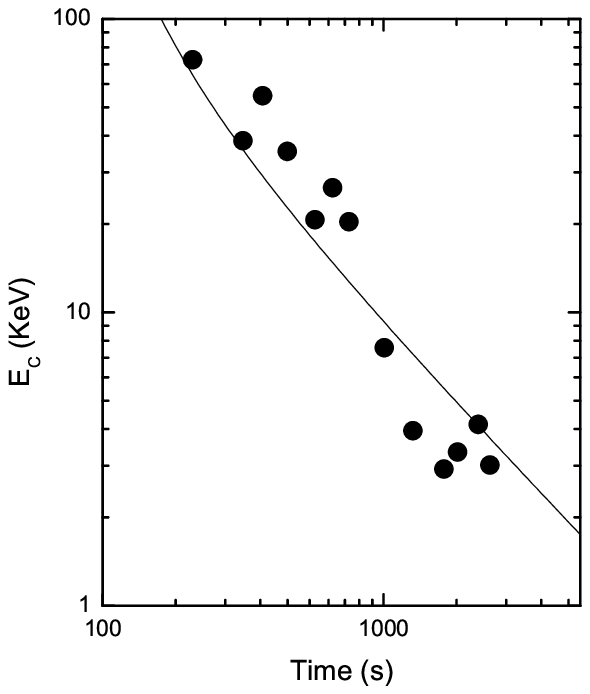}
\\ 
\\
\caption{Testing the third
empirical model with the broad band data of GRB 060218. {\em Left}:
Comparing the third empirical model prediction (solid lines) with the XRT lightcurve and the spectral evolution derived with the XRT data; {\em Right}: Comparing the third empirical model prediction (solid line) with the BAT/XRT joint-fit $E_c$ evolution
(circles, from Ghisellini et al. 2006, following Campana et al. 2006).}
\end{figure}

\clearpage

\renewcommand\thefigure{5}
\begin{figure}
\includegraphics[angle=0]{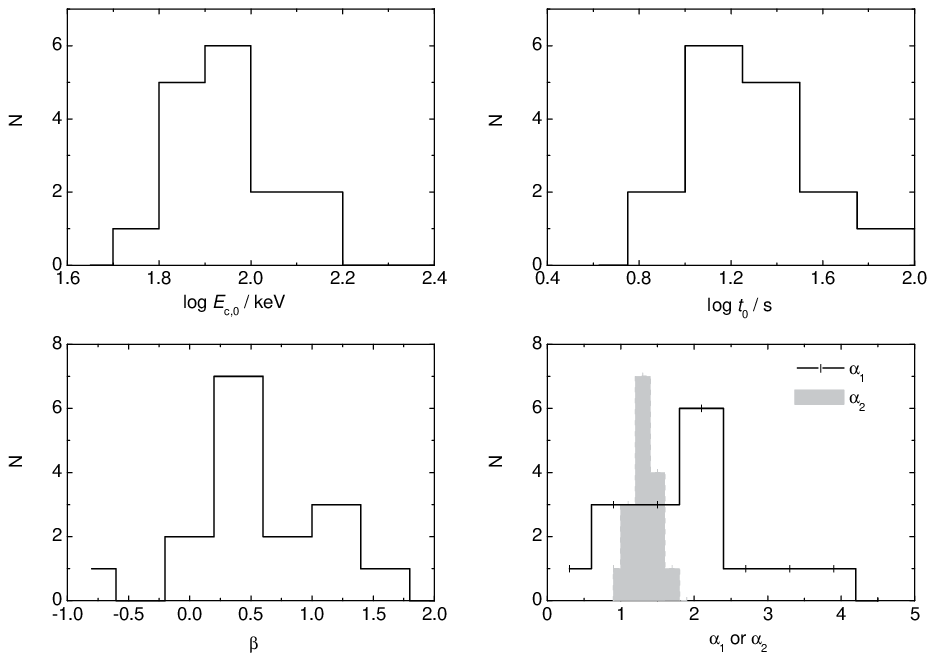}
\caption{Distributions of the model fitting parameters.}
\end{figure}

\clearpage

\renewcommand\thefigure{6}

\begin{figure}
\includegraphics[angle=0]{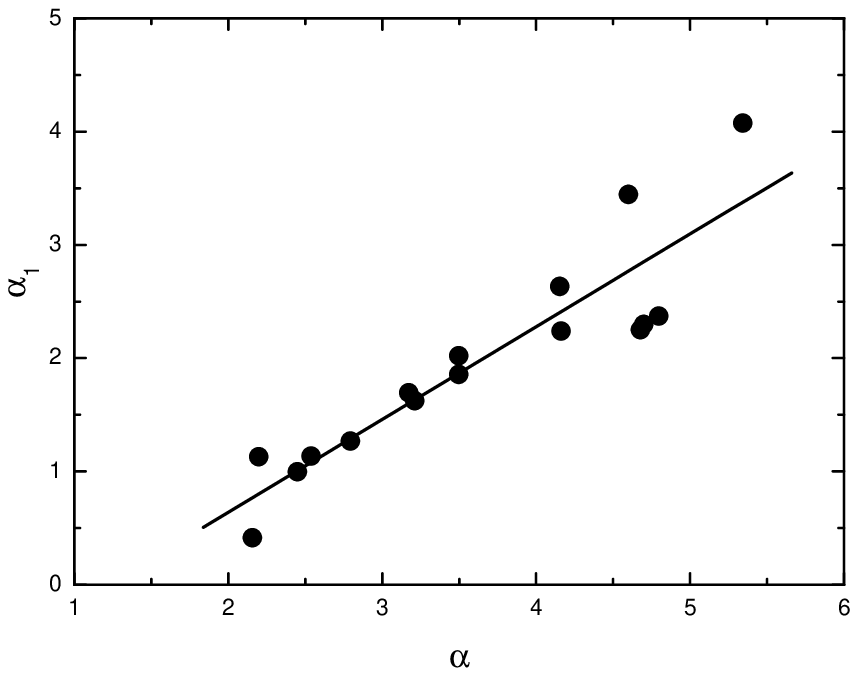}%
\caption{A correlation between the observed tail decay slope $\alpha$ and the decay slope
($\alpha_1$) of the ``spectral amplitude'' (defined in eq.[5]) for the 16 Group B bursts
presented in Figure 2. The solid line is the regression line.}
\end{figure}

\end{document}